\documentclass[preprint,preprintnumbers,showpacs,aps,prd,amssymb]{revtex4-1}

\usepackage{graphicx}
\usepackage{bm}
\usepackage{amsmath}
\usepackage{xcolor}

\def\calH{{\cal H}}

\def\calN{{\cal N}}
\def\calO{{\cal O}}

\def\Bbar{{\bar B}}

\def\hbar{{\bar h}}

\def\sbar{{\bar s}}
\def\ubar{{\bar u}}
\def\nubar{{\bar\nu}}

\def\mhat{{\hat m}}
\def\shat{{\hat s}}
\def\uhat{{\hat u}}

\def\SM{{\rm SM}}
\def\exp{{\rm exp}}
\def\NP{{\rm NP}}
\def\GeV{{\rm GeV}}
\def\TeV{{\rm TeV}}
\def\Br{{\rm Br}}
\def\eff{{\rm eff}}
\def\min{{\rm min}}
\def\dof{{\rm d.o.f.}}
\def\best{{\rm best}}
\def\Re{{\rm Re}}

\def\nn{\nonumber}

\def\Ds{{D^{(*)}}}
\def\Ks{{K^{(*)}}}
\def\Lambdas{{\Lambda^{(*)}}}

\def\Bs2mumu{{B_s\to\mu^+\mu^-}}
\def\B2Knunu{{B^+\to K^+\nu\nubar}}
\def\P5p{{P_5'(B^+\to K^{*+}\mu^+\mu^-)}}

\begin{document}
\title{$\Lambda_b\to\Lambdas\nu\nubar$ and $b\to s$ $B$ decays}
\author{Jong-Phil Lee}
\email{jongphil7@gmail.com}
\affiliation{Sang-Huh College,
Konkuk University, Seoul 05029, Korea}

\begin{abstract}
The baryonic $b\to s$ transition $\Lambda_b\to\Lambda^{(*)}\nu{\bar\nu}$ is analyzed.
We combine the mesonic counterpart $B^+\to K^+\nu{\bar\nu}$ and $B^0\to K^{*0}\nu{\bar\nu}$
as well as other observables involving $B$ mesons like $R(K^{(*)})$, ${\rm Br}(B_s\to\mu^+\mu^-)$,
${\rm Br}(B^+\to K^+\mu^+\mu^-)$, and $P_5'(B^+\to K^{*+}\mu^+\mu^-)$.
%
%
%
%
%
Using the constraints from mesonic sector, we present predictions for the 
currently unobserved baryonic decay modes,
which is very complementary to mesonic modes for probing new physics. 
We find that the new physics scale $M_{\rm NP}$ to be 
$2.04~{\rm TeV}\le M_{\rm NP} \le 11.76~{\rm TeV}$ (at $1\sigma$) for ordinary heavy new mediators.
Our predictions for the branching ratios ${\rm Br}(\Lambda_b\to\Lambda^{(*)}\nu{\bar\nu})$ are
$2.07 (1.07)$ times the standard model estimations,
which could be verified at future colliders. 
%
%
%
%
%
We also find a sum rule for ${\rm Br}(\Lambda_b\to\Lambda\nu{\bar\nu})$ and ${\rm Br}(B\to K^{(*)}\nu{\bar\nu})$
that is very similar to that for $b\to c$ semi-leptonic decays. 
\end{abstract}

\maketitle
\section{Introduction}
%
Processes of $b\to s$ transition are typical cases of the flavor changing neutral currents (FCNCs) 
which are very important in particle physics.
In the standard model (SM) FCNC occurs at loop level and the involved CKM factors are very suppressed.
As a result, FCNC is very sensitive to new physics (NP) effects.
\par
By now a tension with the SM caused by the ratio of the branching ratios \cite{LHCb1705,LHCb2103,Geng2103}
\begin{equation}
R(\Ks)\equiv\frac{\Br(B\to\Ks\mu^+\mu^-)}{\Br(B\to\Ks e^+e^-)}~,
\end{equation}
was resolved by new experimental data from the LHCb \cite{LHCb2212_52,LHCb2212_53}
\begin{eqnarray}
\label{RKsLHCb}
R(K)_L &=& 0.994^{+0.094}_{-0.087}~,~~~
R(K)_C = 0.949^{+0.048}_{-0.047}~,\\
R(K^*)_L &=& 0.927^{+0.099}_{-0.093}~,~~~
R(K^*)_C = 1.027^{+0.077}_{-0.073}~,
\end{eqnarray}
where $L(C)$ means the low (central) $q^2$ region, $[0.1,1.1]~ \GeV^2$ ($[1.1,6.0] ~\GeV^2$).
New experiments are consistent with the SM predictions,
\cite{Hiller0310,Bobeth0709,Geng1704,Bordone1605}
\begin{eqnarray}
R(K)_\SM[1.0,6.0] &=&1.0004^{+0.0008}_{-0.0007}~, \nn\\
R(K^*)_\SM[0.1,1.1] &=& 0.983\pm 0.014~, \nn\\
R(K^*)_\SM[1.1,6.0] &=& 0.996^{+0.002}_{-0.002}~.
\label{RKs_SM}
\end{eqnarray}
Another experiment by the CMS collaboration for $R(K)_C$ \cite{CMS2023}
\begin{equation}
R(K)_C=0.78^{+0.47}_{-0.23}~,
\end{equation}
is also compatible with the SM.
\par
A related process is $B_s\to\mu^+\mu^-$ for which the SM is also successful. 
Experimentally the branching ratio (world average) is \cite{PDG2024}
\begin{equation}
\Br(\Bs2mumu)_\exp = (3.34\pm0.27)\times 10^{-9}~,
\end{equation}
and the SM prediction is \cite{Czaja2407}
\begin{equation}
\Br(\Bs2mumu)_\SM = (3.64\pm0.12)\times 10^{-9}~,
\end{equation}
which is in good agreement with experiments.
\par
But some processes are still in tension with the SM \cite{Capdevila2309}.
For example, experimental measurements of the branching ratio $\Br(B^+\to K^+\mu^+\mu^-)$ 
in some $q^2$ bins \cite{LHCb1403}
\begin{eqnarray}
\Br(B^+\to K^+\mu^+\mu^-)_{\rm LHCb}^{[1.1,2.0]} &=& (0.21\pm0.02)\times 10^{-7}~,\nn\\
\Br(B^+\to K^+\mu^+\mu^-)_{\rm LHCb}^{[4.0,5.0]} &=& (0.22\pm0.02)\times 10^{-7}~,\nn\\
\Br(B^+\to K^+\mu^+\mu^-)_{\rm LHCb}^{[5.0,6.0]} &=& (0.23\pm0.02)\times 10^{-7}~,
\label{B2KmumuLHCb}
\end{eqnarray}
show deviations of $\calO(4\sigma)$ from the SM calculations \cite{Alguero2304} 
\begin{eqnarray}
\Br(B^+\to K^+\mu^+\mu^-)_\SM^{[1.1,2.0]} &=& (0.33\pm0.03)\times 10^{-7}~,\nn\\
\Br(B^+\to K^+\mu^+\mu^-)_\SM^{[4.0,5.0]} &=& (0.37\pm0.03)\times 10^{-7}~,\nn\\
\Br(B^+\to K^+\mu^+\mu^-)_\SM^{[5.0,6.0]} &=& (0.37\pm0.03)\times 10^{-7}~.
\label{B2KmumuSM}
\end{eqnarray}
Also the angular observable $P_5'$ for $B^+\to K^{*+}\mu^+\mu^-$ is measured to be \cite{LHCb2012}
\begin{eqnarray}
\P5p_{\rm LHCb}^{[6.0, 8.0]} &=& -0.15\pm0.41~,\nn\\
\P5p_{\rm LHCb}^{[15.0, 19.0]} &=& -0.24\pm0.17~
\end{eqnarray}
while the SM predicts \cite{Alguero2304}
\begin{eqnarray}
\P5p_\SM^{[6.0, 8.0]} &=& -0.81\pm0.07~,\nn\\
\P5p_\SM^{[15.0, 19.0]} &=& -0.57\pm0.05~.
\end{eqnarray}
\par
Other processes like $B_s\to\phi\mu^+\mu^-$ also possess some tension with the SM,
but we do not consider it in this analysis because of large uncertainties in the local form factors
\cite{Capdevila2309,Gubernari2206}.
%
%
%
%
%
With regard to the selection of observables we adopt two criteria:
(i) theoretical predictions do not suffer from uncontrollable uncertainties; 
(ii) experimental data are precise enough to constrain the relevant Wilson coefficients.
$\Br(B_s\to\phi\mu^+\mu^-)$ is excluded by (i) because it involves large form-factor dependences.
%
%
%
%
%
As pointed out in \cite{Capdevila2309,Alguero2304}, 
total branching ratio and angular observables of $B_s\to\phi\mu^+\mu^-$
suffer from the dependence on the choice of the form factors used.
For example, the use of $B$-meson light-cone sum rules beyond leading twist
(so-called GKvD \cite{GKvD}) alleviates the tension of $\Br(B_s\to\phi\mu^+\mu^-)$
with respect to the SM prediction up to $0.9\sigma$. 
%
%
%
%
%
On the other hand, observables involving $K^*$ such as $R(K^*)$ and $\P5p$ are
defined by fractions where hadronic uncertainties would be reduced, though not completely cancelled.
%
%
%
%
%
Also we do not consider $R(K_S)$ by (ii)
because experimental data
$R(K_S)=0.66^{+0.20}_{-0.14}({\rm stat})^{+0.02}_{-0.04}({\rm sys})$ show
still large uncertainties compared to $R(\Ks)$ \cite{LHCb2110}.
%
%
%
%
%
Moreover, it probes the same Wilson coefficient combinations as $R(K)$ which has smaller uncertainties 
and is already included.
In case of $P_5^\prime$, there are large uncertainties in some bins. 
However, it involves nontrivial combination of the Wilson coefficients $C_9$ and $C_{10}$ 
through the angular distribution providing irreplaceable information, 
which is not independently constrained by  the branching ratios.
On the other hand, $R(K_S)$ is a simple fraction of the branching ratios,
and $R(K)$ gives similar information on the same combination of the Wilson coefficients with more precision.
Thus large uncertainties in $R(K_S)$ reduces its validity relative to $P_5^\prime$.
%
%
%
%
%
\par
In addition, recent measurements of $B\to K\nu\nu$ branching ratios at Belle II reveal \cite{Belle2_2311}
\begin{equation}
\Br(\B2Knunu)_\exp = (2.3\pm 0.7)\times 10^{-5}~,
\end{equation}
while the SM calculation \cite{Becirevic2301}
\begin{equation}
\Br(\B2Knunu)_\SM = (4.43\pm0.31)\times 10^{-6}~,
\end{equation}
predicts much smaller branching ratio.
On the other hand, for $B^0\to K^{*0}\nu\nubar$, the Belle collaboration search results in  \cite{Belle1702}
\begin{equation}
\Br(B^0\to K^{*0}\nu\nubar) < 1.8\times 10^{-5}~,
\label{ctr_Ks}
\end{equation}
which is compatible with the SM estimation \cite{Becirevic2301}
\begin{equation}
\Br(B^0\to K^{*0}\nu\nubar)_\SM = (9.47\pm1.40)\times 10^{-6}~.
\end{equation}
Theoretical studies on $B\to\Ks\nu\nubar$ can be found in 
\cite{Bause2309,Allwicher2309,Chen2401}.
Combined analysis with other $B$ decays indicates that non-zero coupling to $\tau$ flavor is crucial.
\par
At this stage, it would be very interesting to check whether the tension also appears in the baryonic sector.
For example, observed branching ratio of $\Lambda_b\to\Lambda\mu^+\mu^-$ 
\cite{LHCb1503, LHCb1808} is lower than the SM calculations \cite{Gutsche1301,Boer1410,Das1802}.
The situation seems quite similar to $B^+\to K^+\mu^+\mu^-$ case, as shown 
in Eqs.\ (\ref{B2KmumuLHCb}) and (\ref{B2KmumuSM}).
Recently, $\Lambda_b\to\Lambdas\nu\nubar$ is studied in \cite{Das2507}.
The process is complementary to the mesonic version.
%
%
%
%
%
For example, in the baryonic sector we have different spin structure of $1/2 \to 1/2 ({\rm or}~ 3/2)$ transition
and thus different Lorentz structure.
Unlike $B$ meson, $\Lambda_b$ can be polarized and have more than 10 angular observables.
Compared to mesonic modes, more than about 10 angular observables are available in $\Lambda_b$ decays.
In particular, a forward-backward asymmetry proportional to the longitudinal polarization of $\Lambda_b$ is
sensitive to the NP's chirality structure \cite{Altmannshofer2501}.
In the baryonic decays, there are also hadronic uncertainties but in different ways from the mesonic modes.
Thus possible tensions with respect to the SM in the baryonic mode would imply the robustness of NP.
Meanwhile, $\Lambda_b\to\Lambda$ form factors can be calculated more reliably by lattice QCD
because $\Lambda$ is stable under strong interactions \cite{Yadav2409}.
In addition, experimentally $\Lambda_b$ is abundantly produced at LHCb which makes $\Lambda_b$ 
decay modes such as $\Lambda_b\to\Lambda\mu^+\mu^-$ already accessible at hand.
\cite{LHCb1111,LHCb1405}.
The advantage of baryonic mode is complementarity rather than superiority to mesonic one. 
As will be seen later, a sum rule that connects mesonic and baryonic sectors appears in our analysis,
which is very similar to that for $b\to c$ semi-leptonic decays. 
With the results from the $b\to s$ mesonic sector,
our main concern in this paper is to investigate possible NP effects in 
$\Lambda_b\to\Lambdas\nu\nubar$ decays and predict the branching ratios.
Combined observables include $R(\Ks)$, $\Br(\Bs2mumu)$, $\Br(B^+\to K^+\mu^+\mu^-)$, $\P5p$, and
$\Br(B\to\Ks\nu\nubar)$.
\par
There have been many attempts to explain the $B$ anomalies with specific NP models, such as
supersymmetry (SUSY) \cite{Altmannshofer2002,Bardhan2107,Zheng2410}, 
the leptoquark (LQ) \cite{Hiller1408,Dorsner1603,Bauer1511,Chen1703,Crivellin1703,Calibbi1709,Blanke1801,Nomura2104,Angelescu2103,Du2104,Cheung2204}, 
$Z'$ \cite{Crivellin1501,Crivellin1503,Chiang1706,King1706,Chivukula1706,Cen2104,Davighi2105}, 
new scalars \cite{Hu1612,Crivellin1903,Rose1903,Ho2401}, and
unparticles \cite{JPL2106}etc.
In previous works, we parametrized the relevant Wilson coefficients as $C_j^\NP\sim(v/M_\NP)^\alpha$,
where $v$ is the SM vacuum expectation value and $M_\NP$ is the NP scale 
\cite{JPL2110,JPL2208,JPL2411,JPL2502}.
Here $\alpha$ can be non integers inspired by unparticle scenario.
New heavy mediator contribution corresponds to $\alpha=2$,
and the factor of $(v/M_\NP)^2$ appears when low-energy effective theory (LEET) is matched onto 
SM effective theory (SMEFT) or specific NP models are considered.
In this framework one can easily trace the NP scale $M_\NP$ and also probe possible 
unparticle-like degrees of freedom.
We adopt the same parametrization for the Wilson coefficients involving neutrinos.
\par
The paper is organized as follows.
In Sec.\ II, effective Hamiltonians, Wilson coefficients, matrix elements, and the decay rates are given
for both mesonic and baryonic sectors.
Section III provides our main results and discussions.
The $\chi^2$ fit is implemented and the branching ratios $\Br(\Lambda_b\to\Lambdas\nu\nubar)$ are
predicted.
We conclude in Sec.\ IV.
%
%
%
%
%
\section{Wilson coefficients and the branching ratios}
%
Before analyzing processes involving $\nu\nubar$ in the final states,
let's first consider the case of $b\to s\ell^+\ell^-$.
The $b\to s\ell^+\ell^-$ transition can be described by the following effective Hamiltonian
\begin{equation}
\calH_{\rm eff}(b\to s\ell\ell) = 
-\frac{4G_F}{\sqrt{2}}V_{tb}V_{ts}^*\sum_i \left[
C_i(\mu)\calO_i(\mu)+C'_i(\mu)\calO'_i(\mu)\right]~.
\end{equation} 
Relevant operators for our analysis are \cite{Geng2103,Alonso14,Geng17}
\begin{eqnarray}
\calO_9 &=& \frac{e^2}{16\pi^2}\left(\sbar\gamma^\mu P_L b\right)\left({\bar\ell}\gamma_\mu\ell\right)~,\nn\\
\calO_{10} &=& \frac{e^2}{16\pi^2}\left(\sbar\gamma^\mu P_L b\right)\left({\bar\ell}\gamma_\mu\gamma_5\ell\right)~.
\label{O9O10}
\end{eqnarray}
The primed operators are proportional to the right-handed quarks, 
$\calO'_{9,10}\sim (\sbar\gamma^\mu P_R b)$.
In this analysis we do not consider $\calO'_{9,10}$ operators for simplicity.
\par
The matrix elements for $B\to\Ks$ are given by \cite{Ali99}
\begin{eqnarray}
\langle K(p)|\sbar\gamma_\mu b|B(p_B)\rangle&=&
f_+\left[(p_B+p)_\mu-\frac{m_B^2-m_K^2}{s}q_\mu\right]+\frac{m_B^2-m_K^2}{s}f_0 q_\mu~,\\
\langle K(p)|\sbar\sigma_{\mu\nu} q^\nu(1+\gamma_5)b|B(p_B)\rangle&=&
i\left[(p_B+p)_\mu s -q_\mu(m_B^2-m_K^2)\right]\frac{f_T}{m_B+m_K}~,\\
\langle K^*(p)|(V-A)_\mu|B(p_B)\rangle&=&
-i\epsilon_\mu^*(m_B+m_{K^*})A_1+i(p_B+p)_\mu(\epsilon^*\cdot p_B)\frac{A_2}{m_B+m_{K^*}}\nn\\
&&
+iq_\mu(\epsilon^*\cdot p_B)\frac{2m_{K^*}}{s}(A_3-A_0)
+\frac{\epsilon_{\mu\nu\rho\sigma}\epsilon^{*\nu}p_B^\rho b^\sigma}{m_B+m_{K^*}}2V~,
\end{eqnarray}
where $f_{+,0,T}(s), ~A_{0,1,2}(s), ~T_{1,2,3}(s), V(s)$, and $f_-=(f_0-f_+)(1-\mhat_K^2)/\shat$ are the form factors.
Here,
\begin{eqnarray}
 q&=&p_B-p~,~~~s=q^2=(p_B-p)^2~,\\
 \shat&=&\frac{s}{m_B^2}~,~~~\mhat_i=\frac{m_i}{m_B}~.
 \end{eqnarray}
We adopt the form factors from \cite{Chen2401,Buras1409,FLAG2111,Bharucha1503}.
\par
The differential decay rates for $B\to\Ks\ell^+\ell^-$ with respect to $s$ are given by \cite{Chang2010}
\begin{eqnarray}
\frac{d\Gamma_K}{d\shat}&=&
\frac{G_F^2\alpha^2m_B^5}{2^{10}\pi^5}|V_{tb}V_{ts}^*|^2\uhat_{K,\ell}\left\{
(|A'|^2+|C'|^2)\left(\lambda_K-\frac{\uhat_{K,\ell}^2}{3}\right)
+|C'|^24\mhat_\ell^2(2+2\mhat_K^2-\shat) \right .\nn\\
&&\left. +{\rm Re}(C'D'^*)8\mhat_\ell^2(1-\mhat_K^2)+|D'|^24\mhat_\ell^2\shat\right\}~,
\\
\frac{d\Gamma_{K^*}}{d\shat}&=&
\frac{G_F^2\alpha^2m_B^5}{2^{10}\pi^5}|V_{tb}V_{ts}^*|^2\uhat_{K^*,\ell}\left\{
\frac{|A|^2}{3}\shat\lambda_{K^*}\left(1+\frac{2\mhat_\ell^2}{\shat}\right)
+|E|^2\shat\frac{\uhat_{K^*,\ell}^2}{3}\right.\nn\\
&&
+\frac{|B|^2}{4\mhat_{K^*}^2}\left[\lambda_{K^*}-\frac{\uhat_{K^*,\ell}t^2}{3}+8\mhat^2_{K^*}(\shat+2\mhat_\ell^2)\right]
+\frac{|F|^2}{4\mhat_{K^*}^2}\left[\lambda_{K^*}-\frac{\uhat_{K^*,\ell}^2}{3}+8\mhat_{K^*}^2(\shat-4\mhat_\ell^2)\right]\nn\\
&&
+\frac{\lambda_{K^*}|C|^2}{4\mhat_{K^*}^2}\left(\lambda_{K^*}-\frac{\uhat_{K^*,\ell}^2}{3}\right)
+\frac{\lambda|_{K^*}|G|^2}{4\mhat_{K^*}^2}\left[\lambda_{K^*}-\frac{\uhat_{K^*,\ell}^2}{3}
	+4\mhat_\ell^2(2+2\mhat_{K^*}^2-\shat)\right]\nn\\
&&
-\frac{{\rm Re}(BC^*)}{2\mhat_{K^*}^2}\left(\lambda_{K^*}-\frac{\uhat_{K^*,\ell}^2}{3}\right)(1-\mhat_{K^*}^2-\shat)\nn\\
&&
-\frac{{\rm Re}(FG^*)}{2\mhat_{K^*}^2}\left[\left(\lambda_{K^*}-\frac{\uhat_{K^*,\ell}^2}{3}\right)(1-\mhat_{K^*}^2
	-\shat)-4\mhat_\ell^2\lambda_{K^*}\right]\nn\\
&&\left.
-\frac{2\mhat_\ell^2}{\mhat_{K^*}^2}\lambda_{K^*}\left[{\rm Re}(FH^*)-{\rm Re}(GH^*)(1-\mhat_{K^*}^2)\right]
+\frac{\mhat_\ell^2}{\mhat_{K^*}^2}\shat\lambda_{K^*}|H|^2\right\}~,
\end{eqnarray}
where the kinematic variables are
\begin{eqnarray}
\lambda_H&=&1+\mhat_H^4+\shat^2-2\shat-2\mhat_H^2(1+\shat)~,\\
\uhat_{H,\ell}&=&\sqrt{\lambda_H\left(1-\frac{4\mhat_\ell^2}{\shat}\right)}~.
\end{eqnarray}
Here $A',\cdots, D'$ and $A,\cdots, H$ are the auxiliary functions.
They are defined by the form factors combined with the Wilson coefficients as \cite{Ali99},
%
\begin{eqnarray}
A'&=&C_9 f_+ +\frac{2\mhat_b}{1+\mhat_K}C_7^\eff f_T ~,\\
B'&=&C_9 f_- -\frac{2\mhat_b}{\shat}(1-\mhat_K)C_7^\eff f_T ~,\\
C'&=&C_{10} f_+ ~,\\
D'&=&C_{10} f_- ~,
\end{eqnarray}
and
\begin{eqnarray}
A&=&\frac{2}{1+\mhat_{K^*}}C_9 V+\frac{4\mhat_b}{\shat}C_7^\eff T_1~,\\
B&=&(1+\mhat_{K^*})\left[C_9 A_1+\frac{2\mhat_b}{\shat}(1-\mhat_{K^*})C_7^\eff T_2\right]~,\\ 
C&=&\frac{1}{1-\mhat_{K^*}^2}\left[(1-\mhat_{K^*})C_9 A_2
	+2\mhat_b C_7^\eff \left(T_3+\frac{1-\mhat_{K^*}^2}{\shat}T_2\right)\right]~,\\
D&=&\frac{1}{\shat}\left\{C_9\left[(1+\mhat_{K^*})A_1-(1-\mhat_{K^*})A_2
	-2\mhat_{K^*}A_0\right]-2\mhat_b C_7^\eff T_3\right\}~,\\
E&=&\frac{2}{1+\mhat_{K^*}}C_{10} V ~,\\
F&=&(1+\mhat_{K^*})C_{10} A_1 ~,\\
G&=&\frac{1}{1+\mhat_{K^*}}C_{10} A_2 ~,\\
H&=&\frac{1}{\shat}C_{10}\left[(1+\mhat_{K^*})A_1-(1-\mhat_{K^*})A_2-2\mhat_{K^*} A_0\right] ~.
\end{eqnarray}
%
\par
The angular observable $P_5'$ of $B^+\to K^{*+}\mu^+\mu^-$ is defined by some combinations of
the angular coefficients $J_i^a$ associated with the full differential distribution \cite{Mahmoudi2408}.
\par
Another relevant process of $b\to s\ell^+\ell^-$ transition is $\Bs2mumu$. 
The branching ratio is simply given by
\begin{equation}
\Br(B_s\to\mu^+\mu^-)
= \Br(B_s\to\mu^+\mu^-)_\SM \left|1
	+\frac{C_{10}^{\NP}}{C_{10}^{\SM}}\right|^2~,
\label{Br_th}
\end{equation}
where $C_{10}^{\SM}$ is the SM value of $C_{10}$,
$C_{10}^\SM =-4.41$ \cite{Damir1205,DAlise2403}. 
We will use the Wilson coefficients 
$C_{9,10}^\mu = C_{9,10}^e \equiv C_{9,10}$ by lepton-universal scenario \cite{Wen2305}.
%
%
%
%
%
\par
Now let's move to $b\to s\nu\nubar$ decay.
The relevant effective Hamiltonian is \cite{Allwicher2309,Sumensari2406}
\begin{equation}
\calH_{\rm eff}(b\to s\nu\nubar) =
-\frac{4G_F}{\sqrt{2}}\frac{e^2}{(4\pi)^2}V_{tb}V_{ts}^*
\sum_{k=L,R}\sum_{i,j}C_{k}^{\nu_i\nu_j}\Big[\sbar\gamma_\mu P_k b\Big]
    \Big[{\bar\nu}_i\gamma^\mu(1-\gamma_5)\nu_j\Big] + {\rm h.c.}~.
\end{equation}
Here we only consider the case where $\nu_i=\nu_j$, so the relevant Wilson coefficients are
$C_{L(R)}^{\nu_\ell \nu_\ell}\equiv C_{L(R)}^{\nu_\ell}$ with $\ell = e,~\mu, \tau$.
\par
The differential branching ratios of $B\to\Ks\nu\nubar$ are \cite{Chen2401}
\begin{eqnarray}
&& \frac{d\Br(B^+\to K^+\nu\nubar)}{ds} \nn\\
&=&\sum_\ell\frac{\tau_{B^+}G_F^2\alpha_{em}^2}{768\pi^5m_B^3}|V_{tb}V_{ts}^*|^2
\Big[\lambda(s,m_B^2,m_K^2)\Big]^{3/2} 
\Big[f_+(s)\Big]^2 \left| C_L^{\nu_\ell} + C_R^{\nu_\ell}\right|^2~,
\label{dBrKnunu}
\end{eqnarray}
and 
\begin{eqnarray}
&& \frac{d\Br(B^0\to K^{*0}\nu\nubar)}{ds} \nn\\
&=&\sum_\ell\frac{\tau_{B^0}G_F^2\alpha_{em}^2}{384\pi^5m_B^3}|V_{tb}V_{ts}^*|^2
\Big[\lambda(s,m_B^2,m_{K^*}^2)\Big]^{1/2}
\left(m_B+m_{K^*}\right)^2 s \nn\\
&&\times\left\{\left(
\Big[A_1(s)\Big]^2 + \frac{32 m_B^2 m_{K^*}^2}{s(m_B+m_{K^*})^2}\Big[A_{12}(s)\Big]^2\right)
\left|C_L^{\nu_\ell} - C_R^{\nu_\ell}\right|^2 \right. \nn\\
&&\left. + \frac{\lambda(s, m_B^2, m_{K^*}^2)}{(m_B+m_{K^*})^4}
\Big[V(s)\Big]^2\left|C_L^{\nu_\ell} + C_R^{\nu_\ell}\right|^2\right\} ~,
\label{dBrKsnunu}
\end{eqnarray}
Here $\lambda(x,y,z)=x^2+y^2+z^2-2(xy+yz+zx)$ 
and $A_{12}(s)$ is a linear combination of $A_1(s)$ and $A_2(s)$.
Numerical results for the branching ratios for $B^+\to K^+\nu\nubar$ and $B^0\to K^{*0}\nu\nubar$ 
are \cite{Chen2401} ,
\begin{eqnarray}
\Br(B^+\to K^+\nu\nubar) &=& 
3.46\times 10^{-8} \sum_\ell \left| C_L^{\nu_\ell}+C_R^{\nu_\ell}\right|^2~,\\
\Br(B^0\to K^{*0}\nu\nubar) &=& 
   6.84\times 10^{-8} \sum_\ell \left| C_L^{\nu_\ell}-C_R^{\nu_\ell}\right|^2 
+ 1.36\times 10^{-8} \sum_\ell \left| C_L^{\nu_\ell}+C_R^{\nu_\ell}\right|^2~.
\label{BrKsnunu}
\end{eqnarray}
\par
As for the baryonic version, the relevant matrix elements of $\Lambda_b\to\Lambda$ are \cite{Altmannshofer2501}
\begin{eqnarray}
&&\langle\Lambda(p_1)|\sbar\gamma^\mu b|\Lambda_b(p_0)\rangle \nn\\
&=&\ubar^\Lambda\Bigg\{
  f_T^V(k^2)(m_{\Lambda_b}-m_\Lambda)\frac{k^\mu}{k^2}
+f_\perp^V(k^2)\left[\gamma^\mu
       -\frac{2(m_\Lambda p_0^\mu+m_{\Lambda_b}p_1^\mu}{(m_{\Lambda_b}+m_\Lambda)^2-k^2}\right] \nn\\
&&+f_0^V(k^2)\frac{m_{\Lambda_b}+m_\Lambda}{(m_{\Lambda_b}+m_\Lambda)^2-k^2}
\left[p_0^\mu + p_1^\mu -(m_{\Lambda_b}^2-m_\Lambda^2)\frac{k^\mu}{k^2}\right]
\Bigg\} u^{\Lambda_b}~,\\
&&\langle\Lambda(p_1)|\sbar\gamma^\mu\gamma_5 b|\Lambda_b(p_0)\rangle \nn\\
&=&-\ubar^\Lambda\gamma_5\Bigg\{
  f_T^A(k^2)(m_{\Lambda_b}+m_\Lambda)\frac{k^\mu}{k^2}
+f_\perp^A(k^2)\left[\gamma^\mu
       +\frac{2(m_\Lambda p_0^\mu-m_{\Lambda_b}p_1^\mu}{(m_{\Lambda_b}-m_\Lambda)^2-k^2}\right] \nn\\
&&+f_0^A(k^2)\frac{m_{\Lambda_b}-m_\Lambda}{(m_{\Lambda_b}-m_\Lambda)^2-k^2}
\left[p_0^\mu + p_1^\mu -(m_{\Lambda_b}^2-m_\Lambda^2)\frac{k^\mu}{k^2}\right]
\Bigg\} u^{\Lambda_b}~       
\label{Lb2L}
\end{eqnarray}
where $f_{T,0,\perp}^{V,A}$ are the form factors and $k=p_0-p_1$.
For $\Lambda_b\to\Lambda^*$ transition, we can write 
\begin{equation}
\langle\Lambda^*|\sbar\Gamma b|\Lambda_b\rangle 
=\ubar_\rho\mathcal{G}^\rho\big[\Gamma\big] u^{\Lambda_b}
\label{Lb2Ls}
\end{equation}
for Rarita-Schwinger spinor $u_\rho$.
The involved form factors can be found in \cite{Meinel2009}.
For the branching ratios we adopt the numerical results of \cite{Das2507} 
(with proper adjustments due to different conventions),
\begin{eqnarray}
\Br(\Lambda_b\to\Lambda\nu\nubar) &=& 
  7.84\times 10^{-6}
     - 2.34\times 10^{-7}\sum_\ell\Re\Big[(C_{L,\NP}^{\nu_\ell}-C_{R,\NP}^{\nu_\ell})^*\Big]\times 2 \nn\\
&& -1.62\times 10^{-7}\sum_\ell\Re\Big[(C_{L,\NP}^{\nu_\ell}+C_{R,\NP}^{\nu_\ell})^*\Big]\times2 \nn\\
&& +8.90\times 10^{-9}\sum_\ell\Big|C_{L,\NP}^{\nu_\ell}-C_{R,\NP}^{\nu_\ell}\Big|^2\times 4 \nn\\
&& +6.15\times 10^{-9}\sum_\ell\Big|C_{L,\NP}^{\nu_\ell}+C_{R,\NP}^{\nu_\ell}\Big|^2\times 4~, 
\label{E_BrL}
\end{eqnarray}
and 
\begin{eqnarray}
\Br(\Lambda_b\to\Lambda^*\nu\nubar) &=& 
  3.01\times 10^{-9}
     - 1.31\times 10^{-10}\sum_\ell\Re\Big[(C_{L,\NP}^{\nu_\ell}-C_{R,\NP}^{\nu_\ell})^*\Big]\times 2 \nn\\
&& -2.10\times 10^{-11}\sum_\ell\Re\Big[(C_{L,\NP}^{\nu_\ell}+C_{R,\NP}^{\nu_\ell})^*\Big]\times2 \nn\\
&& +4.99\times 10^{-12}\sum_\ell\Big|C_{L,\NP}^{\nu_\ell}-C_{R,\NP}^{\nu_\ell}\Big|^2\times 4 \nn\\
&& +7.97\times 10^{-13}\sum_\ell\Big|C_{L,\NP}^{\nu_\ell}+C_{R,\NP}^{\nu_\ell}\Big|^2\times 4~.
\label{E_BrLs}
\end{eqnarray}
The branching ratios can be scaled by the SM predictions as
\begin{eqnarray}
R(\Lambda)_{\nu\nu} &\equiv& 
   \frac{\Br(\Lambda_b\to\Lambda\nu\nubar)}{\Br(\Lambda_b\to\Lambda\nu\nubar)_\SM} \nn\\
&=&  
 1  - 2.99\times 10^{-2}\sum_\ell\Re\Big[(C_{L,\NP}^{\nu_\ell}-C_{R,\NP}^{\nu_\ell})^*\Big]\times 2 \nn\\
&& -2.06\times 10^{-2}\sum_\ell\Re\Big[(C_{L,\NP}^{\nu_\ell}+C_{R,\NP}^{\nu_\ell})^*\Big]\times2 \nn\\
&& +1.13\times 10^{-3}\sum_\ell\Big|C_{L,\NP}^{\nu_\ell}-C_{R,\NP}^{\nu_\ell}\Big|^2\times 4 \nn\\
&& +7.83\times 10^{-4}\sum_\ell\Big|C_{L,\NP}^{\nu_\ell}+C_{R,\NP}^{\nu_\ell}\Big|^2\times 4~,    
\label{RLnunu}
\end{eqnarray}
and
\begin{eqnarray}
R(\Lambda^*)_{\nu\nu} &\equiv& 
   \frac{\Br(\Lambda_b\to\Lambda^*\nu\nubar)}{\Br(\Lambda_b\to\Lambda^*\nu\nubar)_\SM} \nn\\
&=&  
 1  - 4.37\times 10^{-2}\sum_\ell\Re\Big[(C_{L,\NP}^{\nu_\ell}-C_{R,\NP}^{\nu_\ell})^*\Big]\times 2 \nn\\
&& -6.98\times 10^{-3}\sum_\ell\Re\Big[(C_{L,\NP}^{\nu_\ell}+C_{R,\NP}^{\nu_\ell})^*\Big]\times2 \nn\\
&& +1.66\times 10^{-3}\sum_\ell\Big|C_{L,\NP}^{\nu_\ell}-C_{R,\NP}^{\nu_\ell}\Big|^2\times 4 \nn\\
&& +2.65\times 10^{-4}\sum_\ell\Big|C_{L,\NP}^{\nu_\ell}+C_{R,\NP}^{\nu_\ell}\Big|^2\times 4~.    
\label{RLsnunu}
\end{eqnarray}
\par
Now we parametrize the Wilson coefficients as discussed in the Introduction:
\begin{eqnarray}
C_{9,10} &=& C_{9,10}^\SM + C_{9,10}^\NP~, ~~~{\rm where} \nn\\
C_{9,10}^\NP &\equiv& \calN  A_{9,10} \left(\frac{v}{M_{\NP}}\right)^\alpha ~,
\label{C_setup1}
\end{eqnarray}
with $\calN=|\alpha_{em}V_{tb}V_{ts}^*|^{-1}$.
Here $M_\NP$ is the NP scale and $A_{9,10}^\ell$ are the involved coefficients, and
$\alpha$ is a free parameter.
$A_{9,10}$ involve the fermionic couplings to NP.
In this framework one can focus explicitly on $\alpha$ and $M_\NP$ in a model-independent way.
The setup encodes the NP contributions caused by heavy tree-level new mediators including unparticles.
Also, the validity of specific NP models can be easily checked.
Involved discussions will appear later.
%
%
%
%
%
Our parametrization is quite phenomenological, inspired by specific NP scenarios and/or 
matching LEET onto SMEFT.
For example, LQ or $Z'$ basically contributes to the Wilson coefficients just like Eq.\ (\ref{C_setup1}).
Though the setup is not based on first principles, it would be practically very useful to probe the NP behavior. 
\par
Similarly, we put
\begin{eqnarray}
C_{L(R)}^{\nu_\ell} &=& C_{L(R),\SM}^{\nu_\ell} + C_{L(R),\NP}^{\nu_\ell}~, ~~~{\rm where} \nn\\
C_{L(R),\NP}^{\nu_\ell} &\equiv& \calN A_{L(R)}^{\nu_\ell} \left(\frac{v}{M_{\NP}}\right)^\alpha~,
\label{C_setup2}
\end{eqnarray}
and $C_{R,\SM}^{\nu_\ell} =0$.
According to \cite{Chen2401} where the SMEFT is considered,
analyses for $R(\Ds)$ and $R(\Ks)$ require non-zero 
$C_{L,\NP}^{\nu_e}=C_{L,\NP}^{\nu_\mu}$, $C_{L,\NP}^{\nu_\tau}$ and $C_{R,\NP}^{\nu_\tau}$.
In what follows we adopt the result and keep only these Wilson coefficients active.
%
%
%
%
%
%
%
%
\section{Results and discussions}
%
Our analysis is based on the $\chi^2$ fitting for $R(\Ks)$, $\Br(\Bs2mumu)$, $\Br(B^+\to K^+\mu^+\mu^-)$, 
$\P5p$, and $\Br(B^+\to K^+\nu\nubar)$, with the constraint of $\Br(B^0\to K^{*0}\nu\nubar)$.
%
%
%
%
%
Through the study of mesonic sector we extract useful information about the relevant Wilson coefficients
for the baryonic decay modes.
Usually in specific NP scenarios new couplings and new heavy masses are involved.
In our parametrization couplings are collectively described by $A_j$ and 
a possible power of NP scale ($\alpha$) are introduced. 
The minimum value of $\chi^2$ per degrees of freedom (d.o.f) is $\chi^2_\min/\dof = 1.79$.
%
%
%
%
%
The value seems somewhat unsatisfactory, but it is found that 
$\Br(B^+\to K^+\mu^+\mu^-)$ in $[0.1, 0.98]$ bin alone provides $\chi^2_\min/\dof\simeq 0.72$.
Without the bin we have $\chi^2_\min/\dof\approx 1$.
Actually the experimental data of $\Br(B^+\to K^+\mu^+\mu^-)_{[0.1, 0.98]}$ shows the smallest Pull value
with respect to the SM prediction compared to other bins \cite{Alguero2304}.
In view of NP it is a very local fluctuation.
Thus the large value of $\chi^2_\min$ does not imply the global and systematic breakdown of our fitting. 
As will be seen later, our results are quite stable when $\Br(B^+\to K^+\mu^+\mu^-)$ in  $[0.1, 0.98]$ bin 
is excluded in the analysis.
%
%
%
%
%
%
\begin{table}
\begin{tabular}{c|cc}
fitting parameters$~~~$ & $~~~$ best-fit values & scan ranges \\\hline\
$\alpha$                   & $~~~ 1.21~~~~$      & $~~~~~ [1.,3.] ~~~~~$  \\
$M_{\rm NP}$ (TeV) & $~~~ 13.17~~~$        & $~~~~~ [1., 50.] ~~~~~$  \\
$~~~A_9~~~$          &  $~~~ -0.089~~~$    & $~~~~~ [-1.,1.] ~~~~~$  \\
$~~~A_{10}~~~$     &  $~~~  0.009~~~$     & $~~~~~ [-1.,1.] ~~~~~$   \\
$~~~A_L^{\nu_e}~~~$   &  $~~~ 0.221~~~~$      & $~~~~~ [-1.,1.] ~~~~~$  \\
$~~~A_L^{\nu_\tau}~~~$  &   $~~~ -0.522~~~$   & $~~~~~ [-1.,1.] ~~~~~$  \\
$~~~A_R^{\nu_\tau}~~~$  &   $~~~  -0.468~~~$   &   $~~~~~[-1., 1.]~~~~~ $ \\
\hline\hline
\end{tabular}
\caption{Best-fit values and scan ranges of fitting parameters.}
\label{T_fit_para}
\end{table}
%
Table \ref{T_fit_para} summarize the fitting parameters and their best-fit values with the scan ranges.
Our best-fit value of $\alpha$ is $\alpha_\best = 1.21$, implying that unparticle-like degrees of freedom
or some new contributions other than ordinary heavy particle mediator might be important.
%
%
%
\begin{table}
\begin{tabular}{c|cccc}
 $q^2$ bins     &   $~~$ experiment   &    SM    &
 $~~$ best-fit $~~$    &     $1\sigma$ ranges \\\hline
 $[0.1, 0.98]~~$   &   $~~0.29\pm0.02$   &   $~~~0.32\pm0.03$   &   $0.23$     &    $[0.20,  0.25]$   \\
 $[1.1, 2.]$           &   $~~0.21\pm0.02$   &   $~~~0.33\pm0,03$   &   $0.23$     &    $[0.21,  0.26]$   \\
 $[2., 3.]$             &   $~~0.28\pm0.02$   &   $~~~0.37\pm0.03$   &   $0.26$     &    $[0.23,  0.28]$   \\
 $[3., 4.]$             &   $~~0.25\pm0.02$   &   $~~~0.37\pm0.03$   &   $0.26$     &    $[0.23,  0.28]$   \\
 $[4., 5.]$             &   $~~0.22\pm0.02$   &   $~~~0.37\pm0.03$   &   $0.26$     &    $[0.23,  0.28]$   \\
 $[5., 6.]$             &   $~~0.23\pm0.02$   &   $~~~0.37\pm0.03$   &   $0.25$     &    $[0.23,  0.28]$   \\
 $[6., 7.]$             &   $~~0.25\pm0.02$   &   $~~~0.37\pm0.03$   &   $0.25$     &    $[0.23,  0.27]$   \\
 $[7., 8.]$             &   $~~0.23\pm0.02$   &   $~~~0.38\pm0.04$   &   $0.25$     &    $[0.22,  0.27]$   \\
 $[15., 22.]$         &   $~~0.85\pm0.05$   &   $~~~1.15\pm0.16$   &   $0.78$     &    $[0.71,  0.86]$   \\        
\hline\hline
\end{tabular}
\caption{Best-fit values and $1\sigma$ ranges of our fittings 
for $\Br(B^+\to K^+ \mu^+\mu^-)\times 10^7$.
The SM predictions and the experimental data are from \cite{Alguero2304,LHCb1403}.}
\label{T_Kmumu}
\end{table}
%
\begin{table}
\begin{tabular}{c|cccc}
 $q^2$ bins  &   $~~$ experiment  &   SM     &    
 $~~~$ best-fit $~~$    &      $1\sigma$ ranges \\\hline
 $[0.1, 0.98]~~$   &   $~~ 0.51\pm0.32$    &     $~~~ 0.68\pm0.14$    &   $0.82$      &    $[0.68,  1.00]$   \\
 $[1.1, 2.5]$         &   $~~ 0.88\pm0.72$    &     $~~~ 0.17\pm0.12$    &   $0.36$      &    $[0.11,  0.77]$   \\
 $[2.5, 4.]$           &   $~~ -0.87\pm1.68$   &     $~~~ -0.45\pm0.11$    &   $-0.14$     &    $[-0.49, 0.50]$   \\
 $[4., 6.]$             &   $~~ -0.25\pm0.41$   &     $~~~ -0.73\pm0.08$    &   $-0.41$     &    $[-0.72,  0.30]$   \\
 $[6., 8.]$             &   $~~ -0.15\pm0.41$   &     $~~~ -0.81\pm0.07$    &   $-0.54$     &    $[-0.79,  0.17]$   \\
 $[15., 19.]$         &   $~~ -0.24\pm0.17$   &     $~~~ -0.57\pm0.05$    &   $-0.46$     &    $[-0.59,  -0.02]$   \\   
\hline\hline
\end{tabular}
\caption{Best-fit values and $1\sigma$ ranges of our fittings 
for $P_5'(B^+\to K^{*+} \mu^+\mu^-)$.
The SM predictions and the experimental data are from \cite{Alguero2304,LHCb2012}.}
\label{T_P5p}
\end{table}
%
\par 
Tables \ref{T_Kmumu} and \ref{T_P5p} show our best-fit values and $1\sigma$ ranges of
$\Br(B^+\to K^+\mu^+\mu^-)$ and $\P5p$ for several $q^2$ bins. 
The $B\to K$ form factors are parameterized as Flavor Lattice Averaging Group (FLAG) \cite{FLAG2111}
and the $B\to K^*$ form factors are taken from the light-cone sum rules \cite{Bharucha1503}.
The best-fit values of $\Br(B^+\to K^+\mu^+\mu^-)$ tend to be lower than the SM predictions, 
while those of $\P5p$ higher than.
%
%
%
\begin{table}
\begin{tabular}{c|ccc}
    & $~~~$ best-fit values &  $~~~ 1\sigma$ ranges \\\hline
 $(R(K)_L-1)\times 10^2$     & $-0.71$    &   $[-0.72, -0.69]$     \\
 $(R(K)_C-1)\times 10^2$     & $0.05$    &   $[0.04, 0.07]$    \\
 $(R(K^*)_L-1)\times 10^2$   & $-1.82$    &   $[-1.83, -1.77]$   \\
 $(R(K^*)_C-1)\times 10^2$   & $-0.50$    &   $[-0.55, -0.41]$    \\
 $~~~\Br(B_s\to\mu\mu)\times 10^9~~~$  & $3.40$    &   $[2.61, 4.37]$   \\
 $~~~\Br(B^+\to K^+\nu\nubar)\times 10^5~~~$           & $1.87$    &   $[0.00, 2.71]$    \\
 $~~~\Br(B^0\to K^{*0}\nu\nubar)\times 10^5~~~$        & $1.19$    &   $[0.01, 1.80]$    \\
 $~~~\Br(\Lambda_b\to\Lambda\nu\nubar)\times 10^5$     &   $1.63$   &   $[0.02, 2.33]$ \\
 $~~~\Br(\Lambda_b\to\Lambda^*\nu\nubar)\times 10^9$  &   $3.23$   &   $[0.07, 5.42]$ \\
 $~~~R(\Lambda)_{\nu\nu}$     &   $2.07$   &   $[0.02, 2.96]$ \\ 
  $~~~R(\Lambda^*)_{\nu\nu}$ &   $1.07$   &   $[0.02, 1.80]$ \\
 $~~~C_9^\NP~~~$   & $ -1.49$       &   $[-3.36, -0.57]$   \\  
 $~~~C_{10}^\NP~~~$    &   $ 0.15$   &  $[-0.42, 0.68]$   \\
 $~~~C_{L,\NP}^{\nu_e}~~~$   &   $3.70$   &   $[-3.92,14.52]$  \\       
 $~~~C_{L,\NP}^{\nu_\tau}~~~$   &   $-8.76$   &   $[-11.47,18.64]$  \\    
 $~~~C_{R,NP}^{\nu_\tau}~~~$   &   $-7.86$   &   $[-13.29,12.51]$  \\    
\hline\hline
\end{tabular}
\caption{Best-fit values and $1\sigma$ range of our fittings 
for observable and the Wilson coefficients.}
\label{T_obs}
\end{table}
%
\par
In Table \ref{T_obs} we provide the best-fit values and $1\sigma$-allowed ranges of observables and
the Wilson coefficients. 
Best-fit values of $R(\Ks)$ are very close to the SM predictions. 
This is mainly due to non-zero $A_{9,10}^e$ (or, equivalently, $C_{9,10}^e$).
\par
One of the main results of our analysis is that the best-fit value of the branching ratio 
$\Br(\Lambda_b\to\Lambda\nu\nubar)$ is higher than the SM predictions, or equivalently,
that of $R(\Lambda)_{\nu\nu}$ is larger than unity,
while $\Br(\Lambda_b\to\Lambda^*\nu\nubar)$ is close to the SM value.
Our results (Table \ref{T_obs}) are consistent with previous work of Ref.\ \cite{Das2507} 
though the best-fit values of $\Br(\Lambda_b\to\Lambda\nu\nubar)$ and $R(\Lambda)_{\nu\nu}$
are larger than the upper bound of \cite{Das2507}.
As a comparison in the mesonic sector, one has \cite{Das2507}
\begin{eqnarray}
R(K)_{\nu\nu} &\equiv& 
\frac{\Br(B^+\to K^+\nu\nubar)_\exp}{\Br(B^+\to K^+\nu\nubar)_\SM} = 5.4\pm 1.6~, \nn\\
R(K^*)_{\nu\nu} &\equiv& 
\frac{\Br(B^0\to K^{*0}\nu\nubar)_\exp}{\Br(B^0\to K^{*0}\nu\nubar)_\SM} < 2.7~.
\end{eqnarray}
In this respect, $R(\Lambda)_{\nu\nu}^\best = 2.07$ might not be a surprise.
The reason for large branching ratios is that $|C_{L,\SM}^{\nu_\ell}|= |-6.32(7)|$ 
\cite{Buras1409,Brod2105,Bause2309}
in Eq.\ (\ref{BrKsnunu}) is not big enough to explain the experimental data of $\Br(B^+\to K^+\nu\nubar)$.
To enhance the branching ratio, 
$|C_{L,\SM}^{\nu_\ell}+C_{L,\NP}^{\nu_\ell}+C_{R,\NP}^{\nu_\ell}|$ should be large
(but with the constraint by Eq.\ (\ref{ctr_Ks})).
As a consequence, $\Br(\Lambda_b\to\Lambda\nu\nubar)$ in Eq.\ (\ref{E_BrL}) gets larger.
By contrast, in $\Br(\Lambda_b\to\Lambda^*\nu\nubar)$ terms including 
$(C_{L,\NP}^{\nu_\ell}+C_{R,\NP}^{\nu_\ell})$ are not dominant as seen in Eq.\ (\ref{E_BrLs}).
As for $\Br(\Lambda_b\to\Lambda^*\nu\nubar)$ and $\Br(B^0\to K^{*0}\nu\nubar)$,
terms of $(C_{L,\NP}^{\nu_\ell}-C_{R,\NP}^{\nu_\ell})$ are also important (or even dominant)
as in Eqs.\ (\ref{BrKsnunu}) and (\ref{E_BrLs}),
but $\Br(B^0\to K^{*0}\nu\nubar)$ is bounded above.
As a consequence $\Br(\Lambda_b\to\Lambda^*\nu\nubar)$ cannot be large enough.
%
%
%
%
%
In the baryonic modes uncertainties of the related form factors also matter.
As discussed in \cite{Altmannshofer2501}, uncertainties in $\Br(\Lambda_b\to\Lambda\nu\nubar)_\SM$
amount to 14\%, about 51\% of which come from the form factors.
Now assuming 7\% of uncertainty coming from the form factors, we find that the resulting fluctuation is
estimated to be $R(\Lambda)_{\nu\nu}^\best = 2.07^{+0.16}_{-0.15}$.
The Pull value with respect to unity is $\approx 6.9$, far beyond the SM prediction.
Therefore one can safely conclude that $R(\Lambda)_{\nu\nu}$ is still very sensitive to NP even when
considering the form factor uncertainties.
%
%
%
%
%
%
\begin{figure}
\begin{tabular}{cc}
\hspace{-1cm}\includegraphics[scale=0.12]{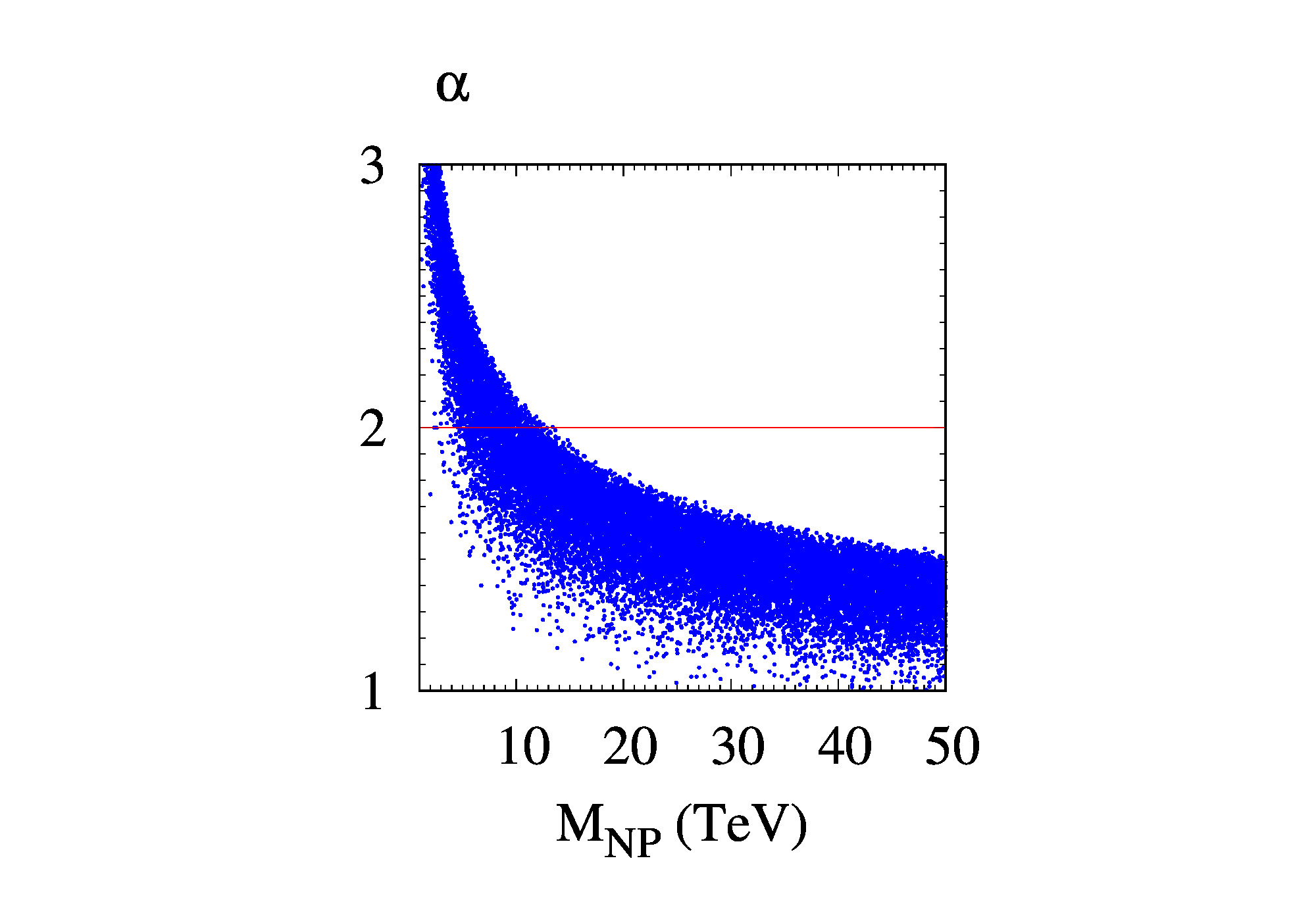} &
\hspace{-1cm}\includegraphics[scale=0.12]{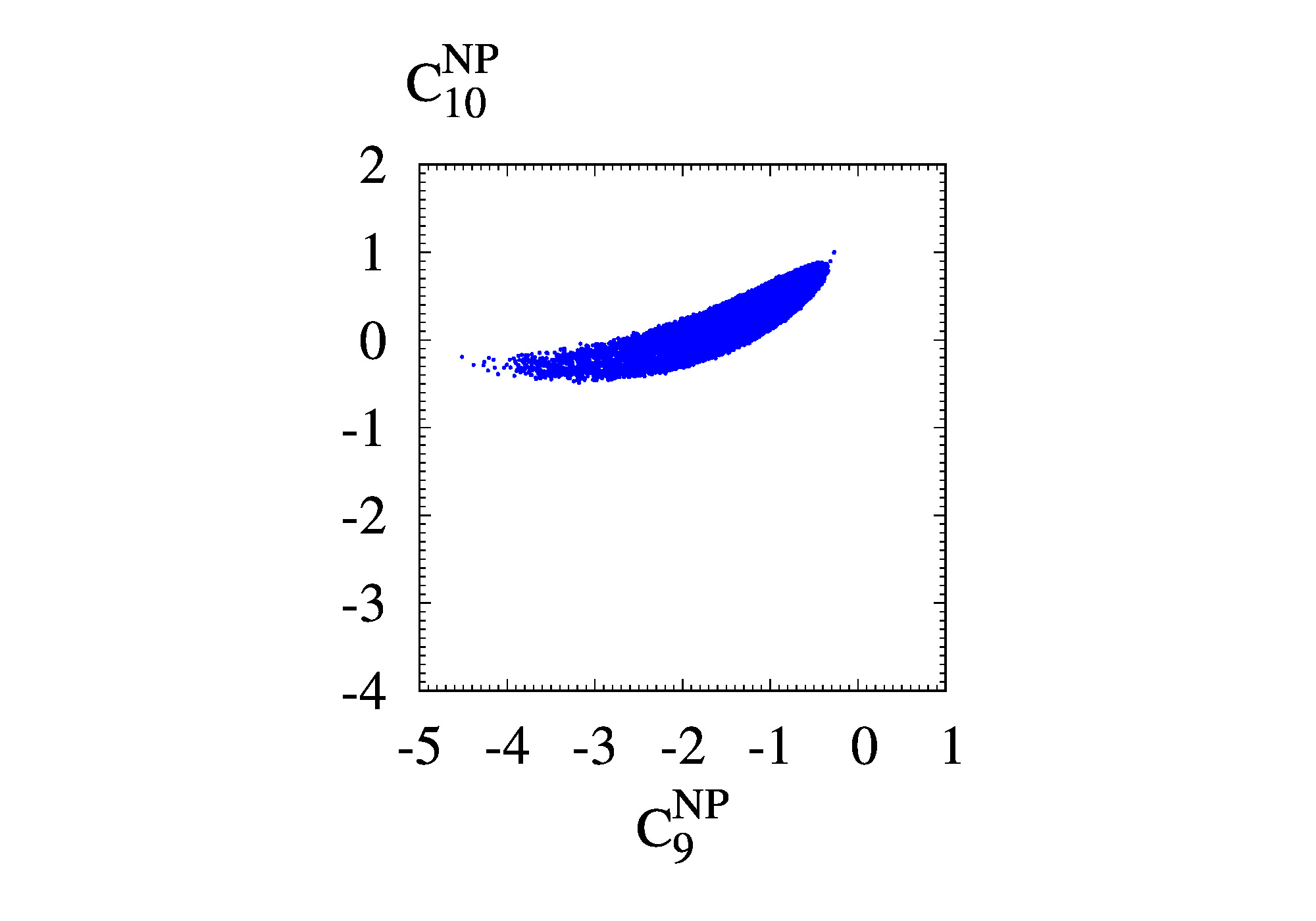} \\
(a) & (b) \\
\hspace{-1cm}\includegraphics[scale=0.12]{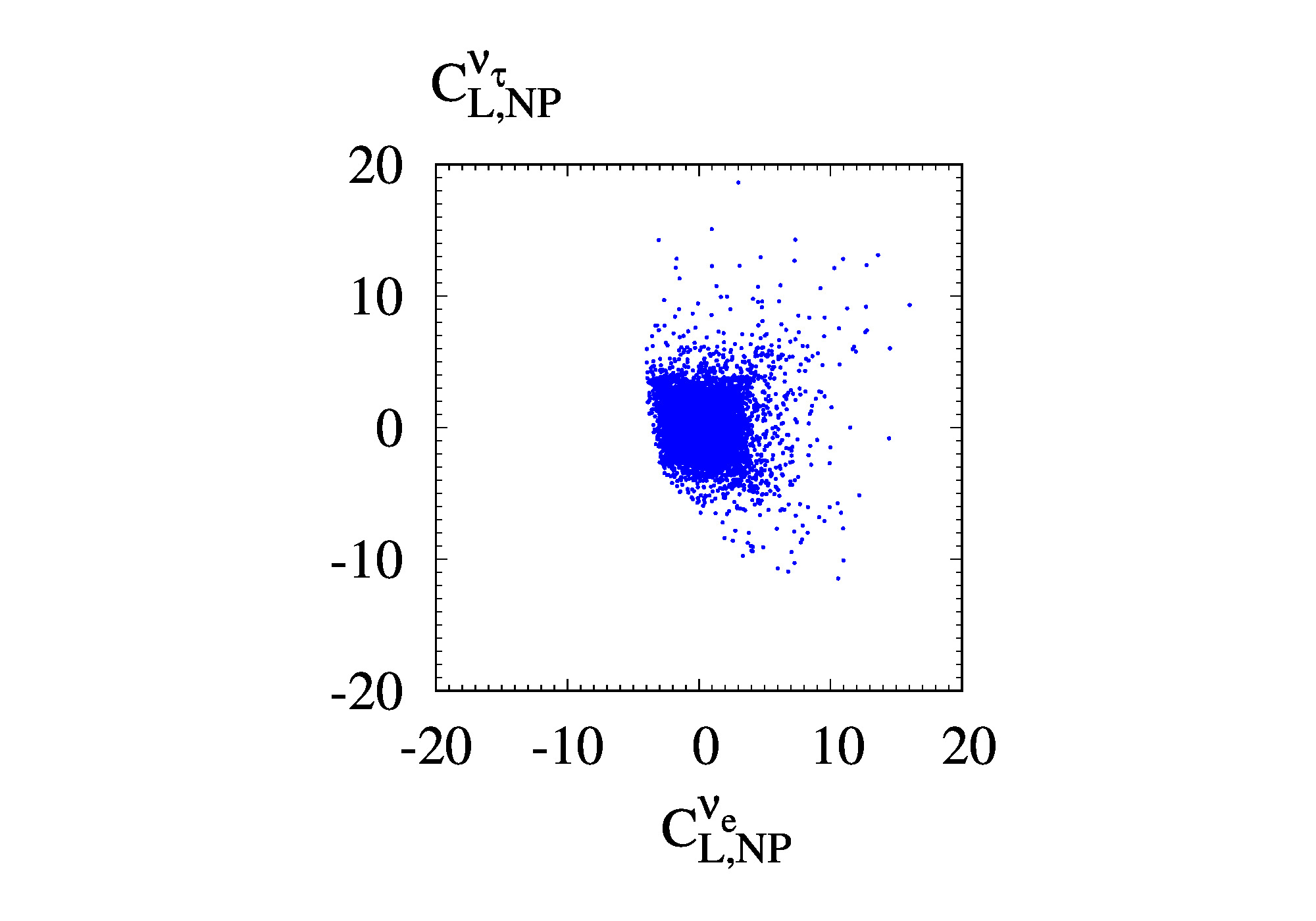} &
\hspace{-1cm}\includegraphics[scale=0.12]{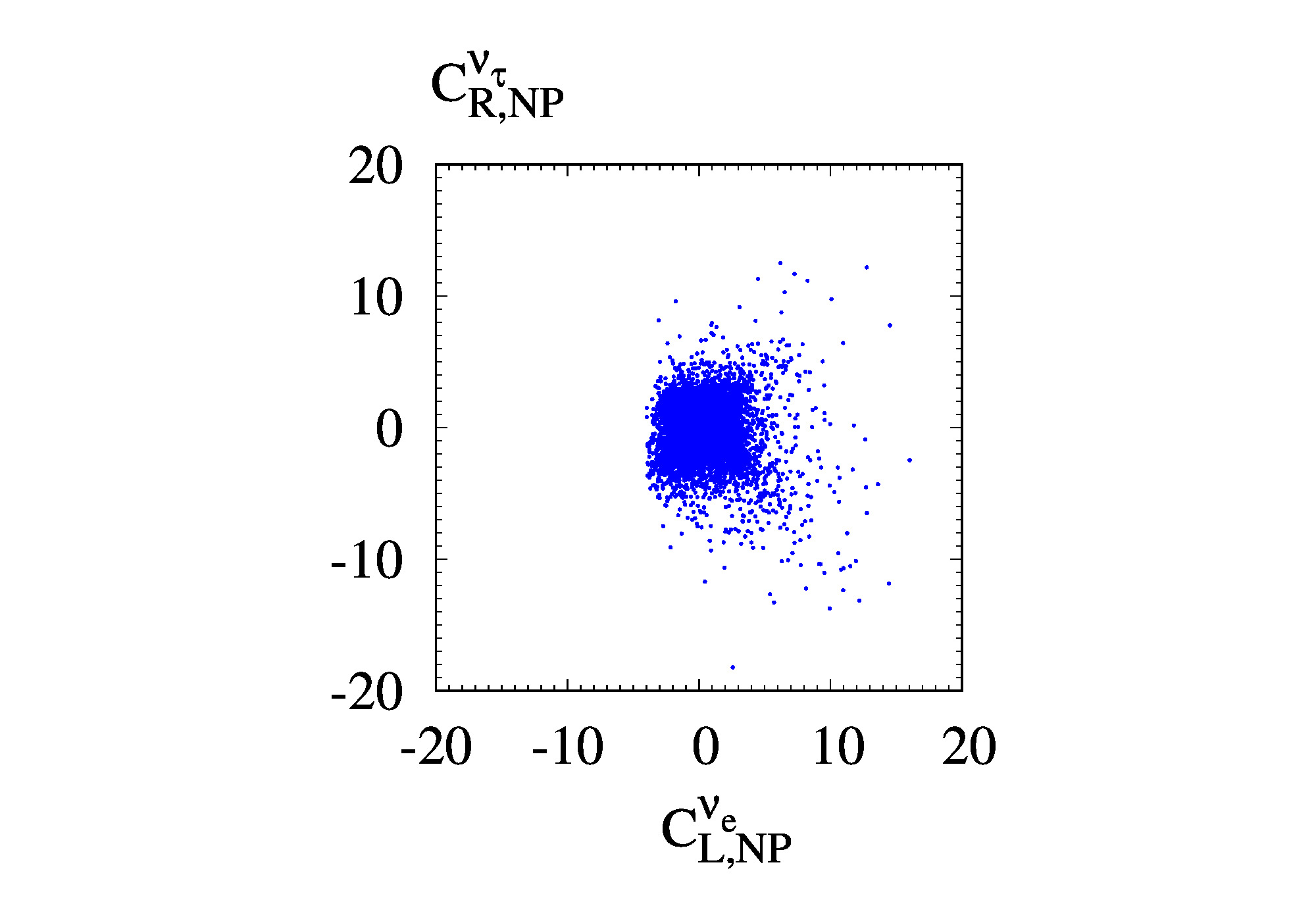} \\
(c) & (d) \\
\hspace{-1cm}\includegraphics[scale=0.12]{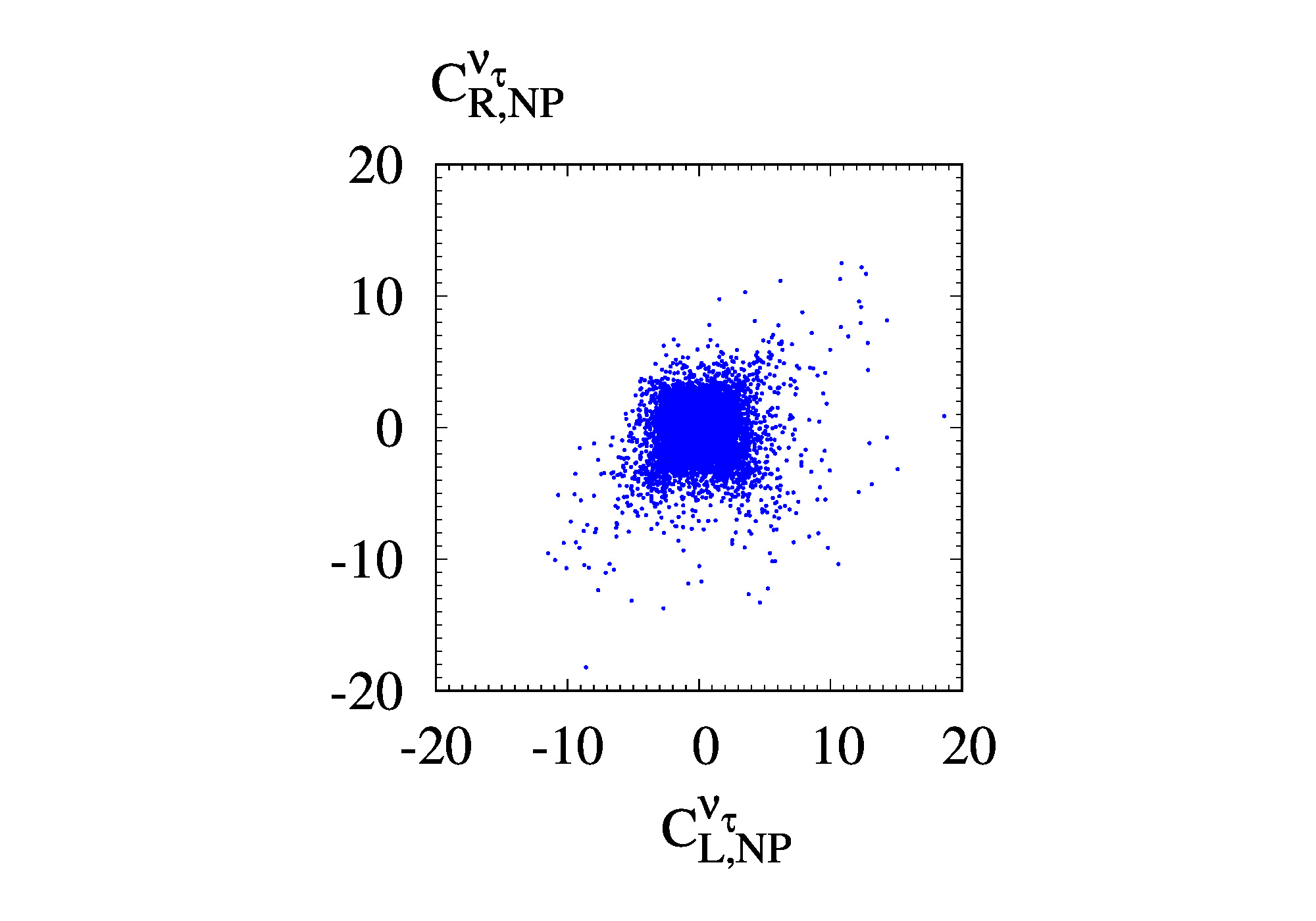} & \\
(e) &
\end{tabular}
\caption{\label{F_para} 
Allowed regions at the $2\sigma$ level for
(a) $\alpha$ vs. $M_\NP$, 
(b) $C_{10}^\NP$ vs. $C_9^\NP$, 
(c) $C_{L,\NP}^{\nu_\tau}$ vs. $C_{L,\NP}^{\nu_e}$, 
(d) $C_{R,\NP}^{\nu_\tau}$ vs. $C_{L,\NP}^{\nu_e}$, and
(e) $C_{R,\NP}^{\nu_\tau}$ vs. $C_{L,\NP}^{\nu_\tau}$.
}
\end{figure}
\par
Figure \ref{F_para} depicts allowed regions of $\alpha$-$M_\NP$ and Wilson coefficients 
at the $2\sigma$ level.
As shown in Fig.\ \ref{F_para} (b), $C_9^\NP$ favors negative values.
It is necessary to fit the experimental data $\Br(B^+\to K^+\mu^+\mu^-)$ and $\P5p$ \cite{JPL2502}.
Figures \ref{F_para} (c)-(e) indicate that $|C_{L(R),\NP}^{\nu_\ell}|$ can be larger than $|C_{9,10}^\NP|$
or $|C_{L,\SM}^{\nu_\ell}|$.
The reason is that, as mentioned before, $|C_{L,\SM}^{\nu_\ell}|$ alone cannot explain large branching ratio
$\Br(B^+\to K^+\nu\nubar)$.
Also, $C_{L,\NP}^{\nu_\ell}$ can be positively large, which is due to the fact that $C_{L,\SM}^{\nu_\ell}$ is negative.
%
%
%
%
%
\begin{figure}
\begin{tabular}{cc}
\hspace{-1cm}\includegraphics[scale=0.12]{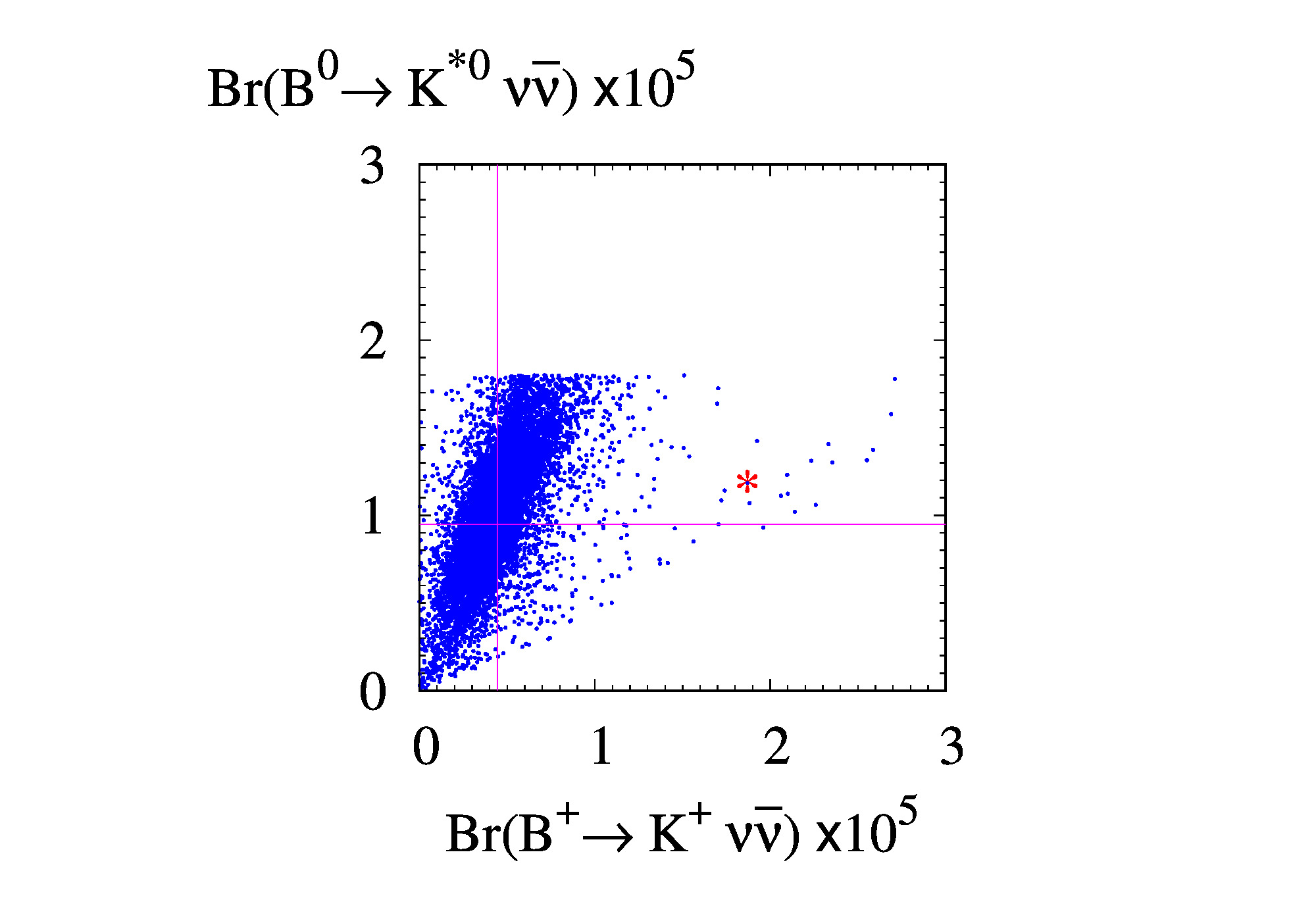} &
\hspace{-1cm}\includegraphics[scale=0.12]{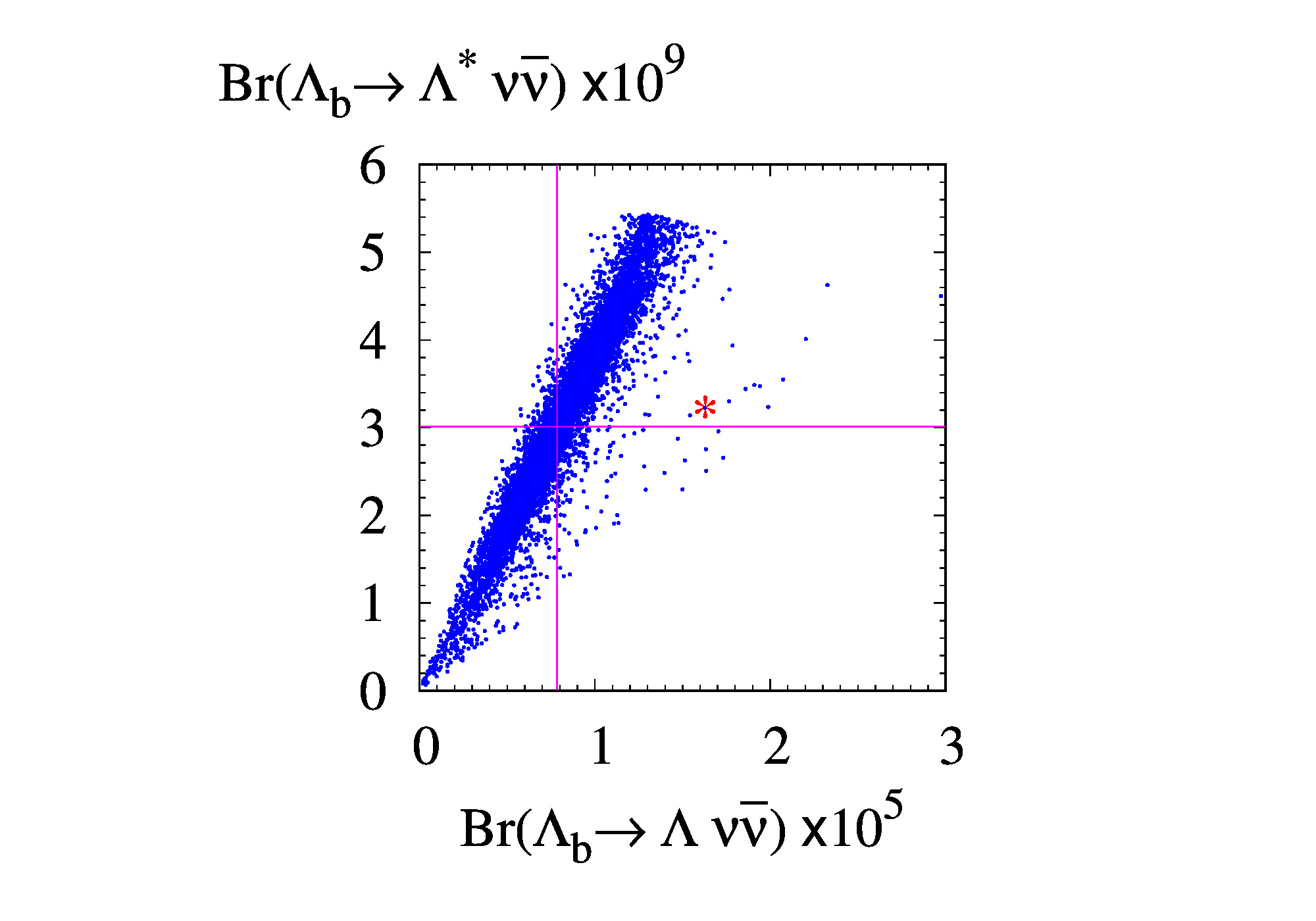} \\
(a) & (b) \\
\hspace{-1cm}\includegraphics[scale=0.12]{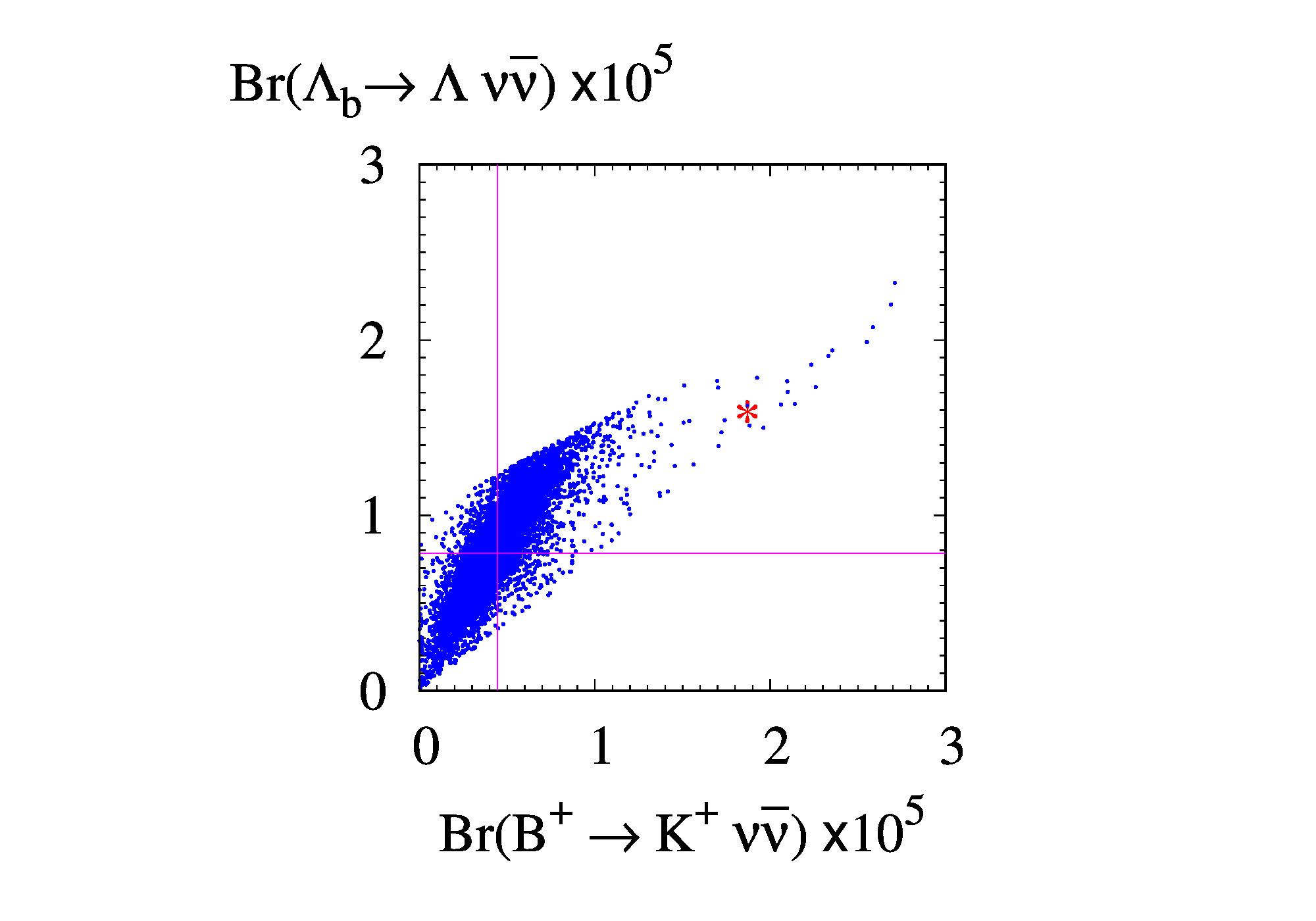} & 
\hspace{-1cm}\includegraphics[scale=0.12]{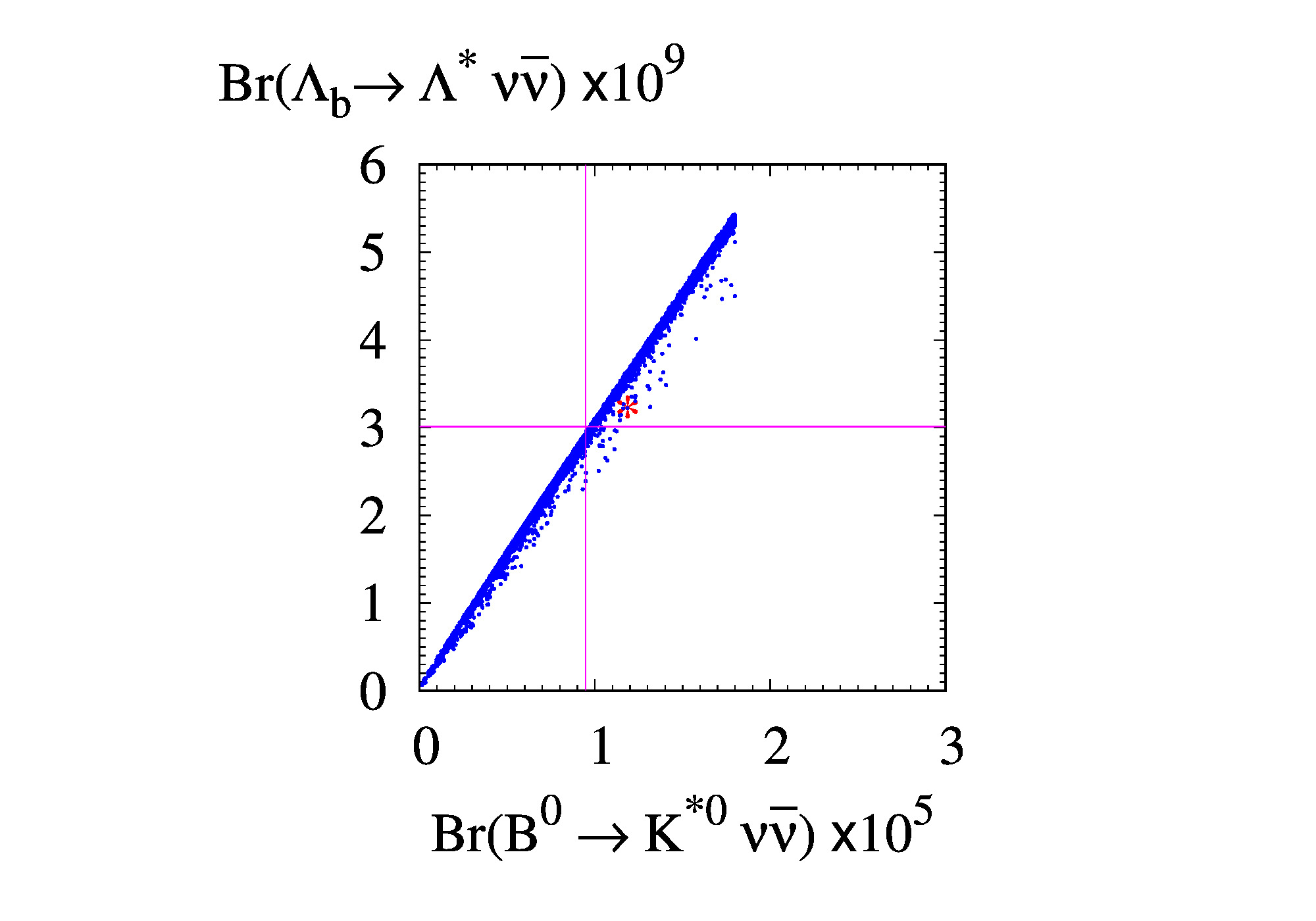} \\
(c) & (d) \\
\hspace{-1cm}\includegraphics[scale=0.12]{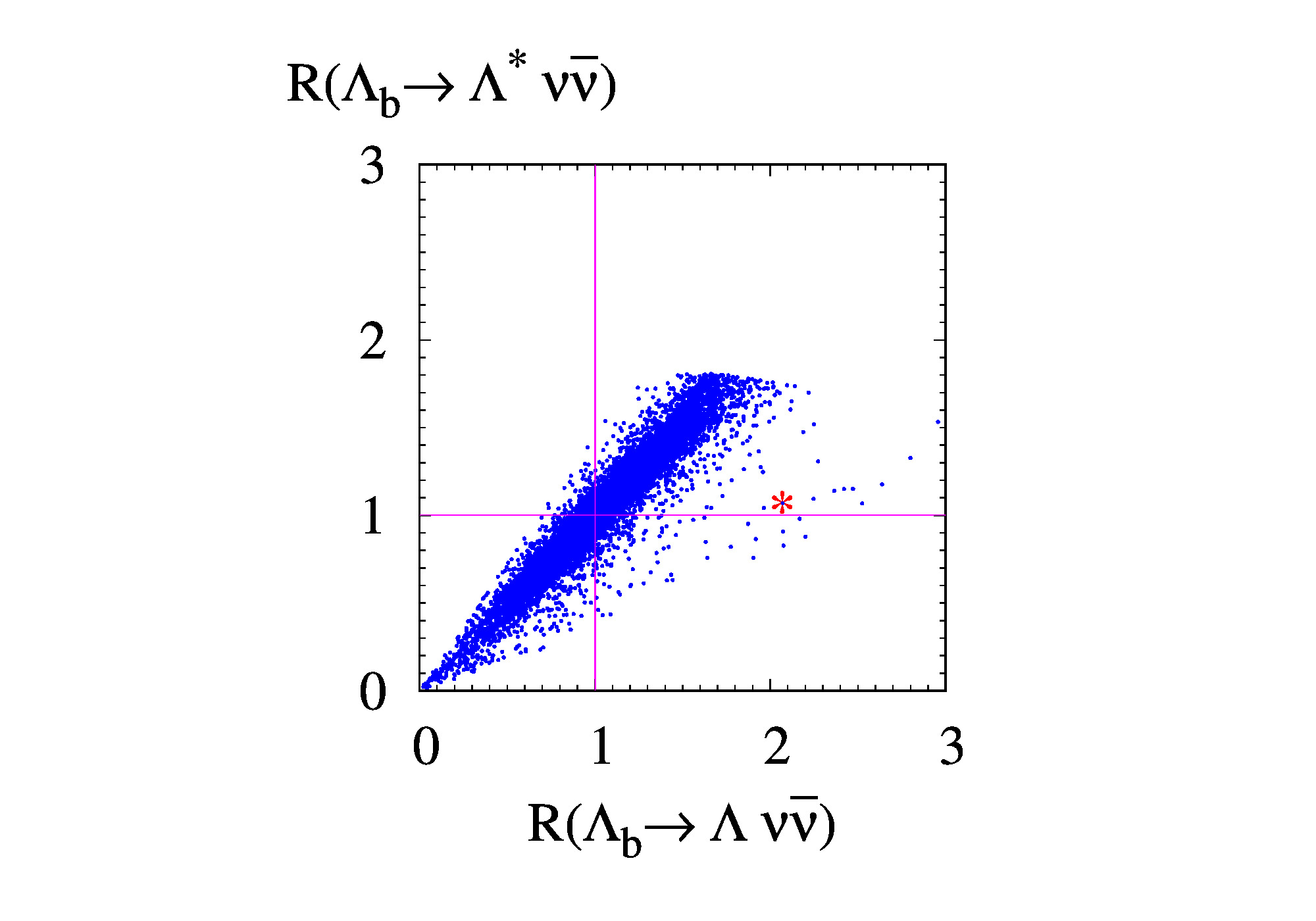} & 
\hspace{-1cm}\includegraphics[scale=0.12]{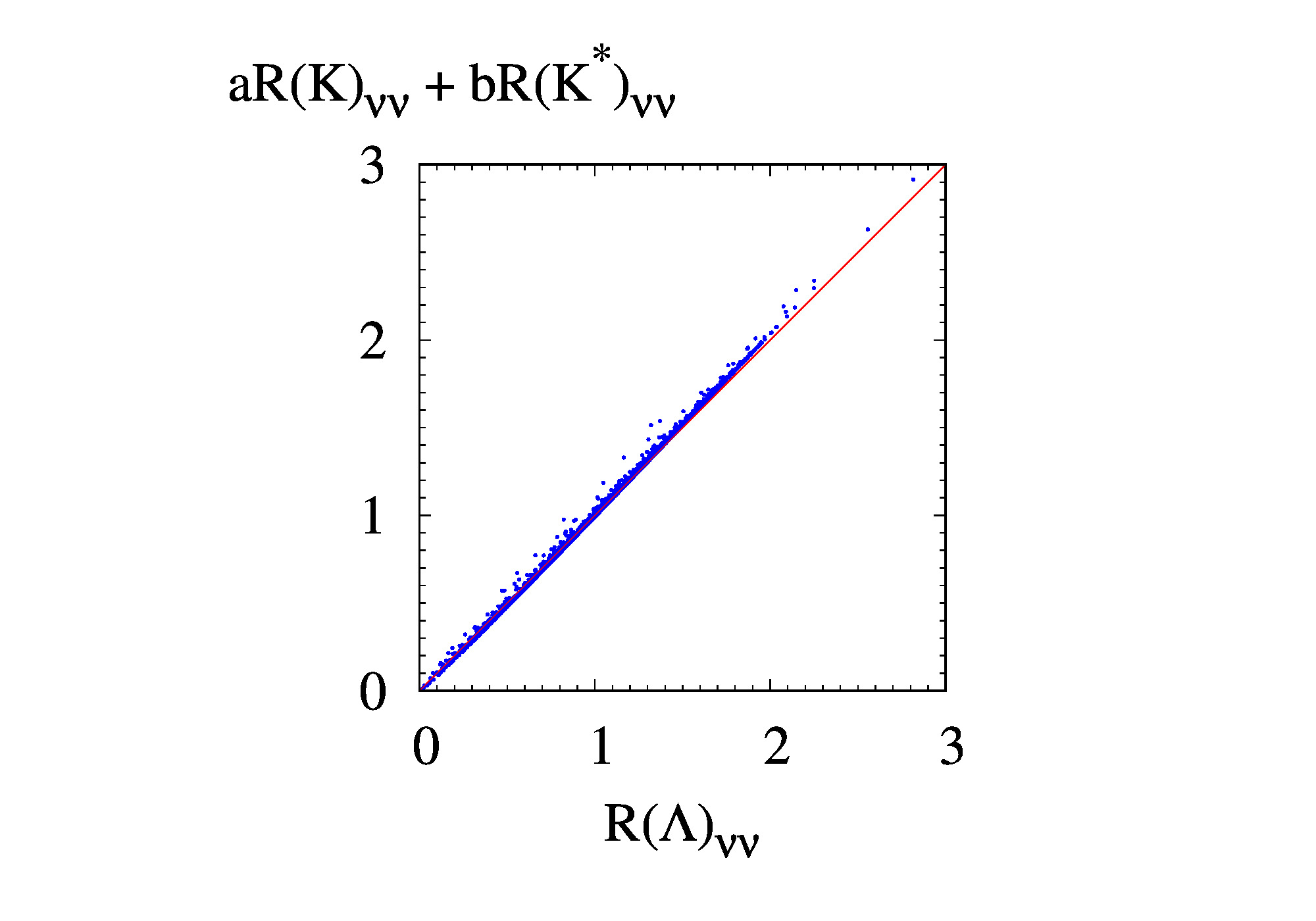} \\
(e) & (f)
\end{tabular}
\caption{\label{F_obs} 
Allowed regions at the $2\sigma$ level for
(a) $\Br(B^0\to K^{*0}\nu\nubar)$ vs. $\Br(B^+\to K^+\nu\nubar)$,
(b) $\Br(\Lambda_b\to\Lambda^*\nu\nubar)$ vs. $\Br(\Lambda_b\to\Lambda\nu\nubar)$,
(c) $\Br(\Lambda_b\to\Lambda\nu\nubar)$ vs. $\Br(B^+\to K^+\nu\nubar)$,
(d) $\Br(\Lambda_b\to\Lambda^*\nu\nubar)$ vs. $\Br(B^0\to K^{*0}\nu\nubar)$, 
(e) $R(\Lambda^*)_{\nu\nu}$ vs. $R(\Lambda)_{\nu\nu}$, and
(f) $a R(K)_{\nu\nu}+ b R(K^*)_{\nu\nu}$ vs. $R(\Lambda)_{\nu\nu}$ with
$(a,b)=(0.30,0.69)$.
Magenta lines are the SM predictions, the red stars are our best-fit points, and
the red line in (f) is $y=x$.
}
\end{figure}
\par
In Fig.\ \ref{F_obs} correlations between various branching ratios are given.
Here the magenta lines are the SM predictions and the red stars are our best-fit values.
Most of plots are located around the SM points rather than the best-fit ones.
This is because $R(\Ks)$ and $\Br(B_s\to\mu^+\mu^-)$ are in good agreement with the SM.
One can see in Fig.\ \ref{F_obs} (d) that best-fit values of 
$\Br(B^0\to K^{*0}\nu\nubar)$ and $\Br(\Lambda_b\to\Lambda^*\nu\nubar)$ are close to the SM values.
As discussed before, $\Br(\Lambda_b\to\Lambda^*\nu\nubar)$ cannot be much larger than the SM predictions.
The branching ratios $\Br(B^0\to K^{*0}\nu\nubar)$ and $\Br(\Lambda_b\to\Lambda^*\nu\nubar)$ share
a common feature where $(C_{L,\NP}^{\nu_\ell}-C_{R,\NP}^{\nu_\ell})$ is dominant over
$(C_{L,\NP}^{\nu_\ell}+C_{R,\NP}^{\nu_\ell})$.
The feature is reflected in Fig.\ \ref{F_obs} (d).
This is not true for $\Br(B^+\to K^+\nu\nubar)$ and $\Br(\Lambda_b\to\Lambda\nu\nubar)$.
%
%
%
%
%
Figure \ref{F_obs} (e) is a replica of Fig.\ \ref{F_obs} (b) scaled by the SM values.
For $b\to c$ transition there is a well-known sum rule of  \cite{Duan2410,Endo2501,Endo2509}
\begin{equation}
\frac{R_{\Lambda_c}}{R_{\Lambda_c}^\SM}\simeq
0.28\frac{R_D}{R_D^\SM} + 0.72\frac{R_{D^*}}{R_{D^*}^\SM}~,
\label{SR1}
\end{equation}
where $R_{H_c}\equiv\Br(H_b\to H_c\tau\nubar_\tau)/\Br(H_b\to H_c\ell\nubar_\ell)$ and
$R_{H_c}^\SM$ is the SM predictions.
One can expect a similar sum rule for $b\to s\nu\nubar$ processes.
Figure \ref{F_obs} (f) is the numerical result from our allowed dataset.
We have
\begin{equation}
R(\Lambda)_{\nu\nu} = a R(K)_{\nu\nu} + b R(K^*)_{\nu\nu}~,
\label{SR2}
\end{equation}
with $(a, b) = (0.30, 0.69)$.
As in Eq.\ (\ref{SR1}), $a+b\approx 1$.
Note that $b\to c\ell\nubar$ and $b\to s\nu\nubar$ decays are quite different.
The former is charged current and dominated by tree-level process.
Also one can expect the heavy quark symmetry for $b$ and $c$ quarks, which underlies the sum rule
\cite{Endo2501,Endo2509}.
In the heavy quark limit the sum rule holds exactly.
On the other hand the latter is FCNC and there is no heavy quark symmetry.
In this regard very similar sum rules in two different channels are quite remarkable.
Since current analysis is not based on specific NP models 
we expect the relation of Eq.\ (\ref{SR2}) to be quite general.
By now there are no measurements of $R(\Lambda, K^*)_{\nu\nu}$.
Assuming the sum rule of Eq.\ (\ref{SR2}) is robust, 
measurements in mesonic sector provide strong indication to baryonic one, and vice versa.
And experimental checks of the sum rule would be quite challenging. 
%
%
%
%
\begin{figure}
\begin{tabular}{cc}
\hspace{-1cm}\includegraphics[scale=0.12]{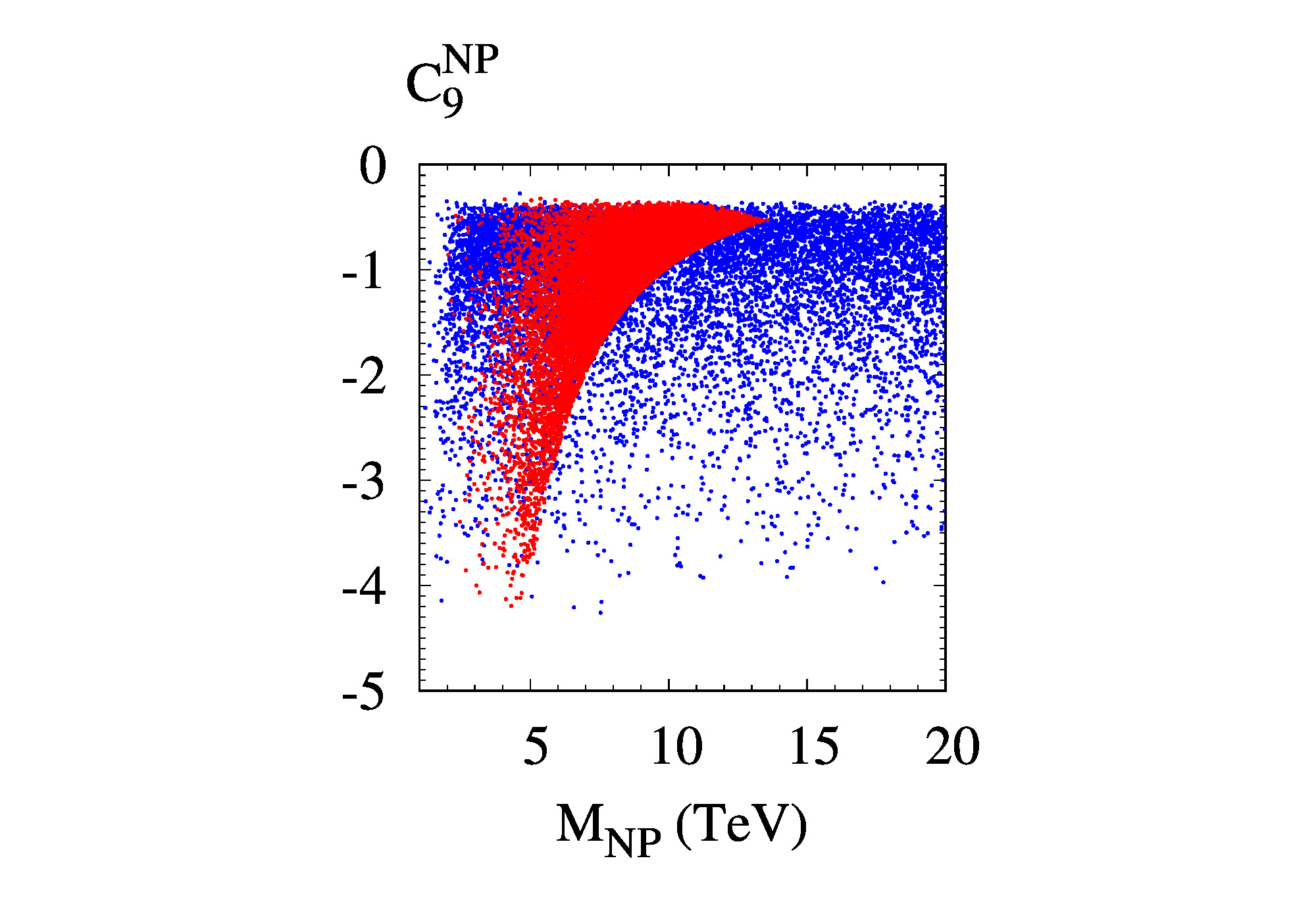} &
\hspace{-1cm}\includegraphics[scale=0.12]{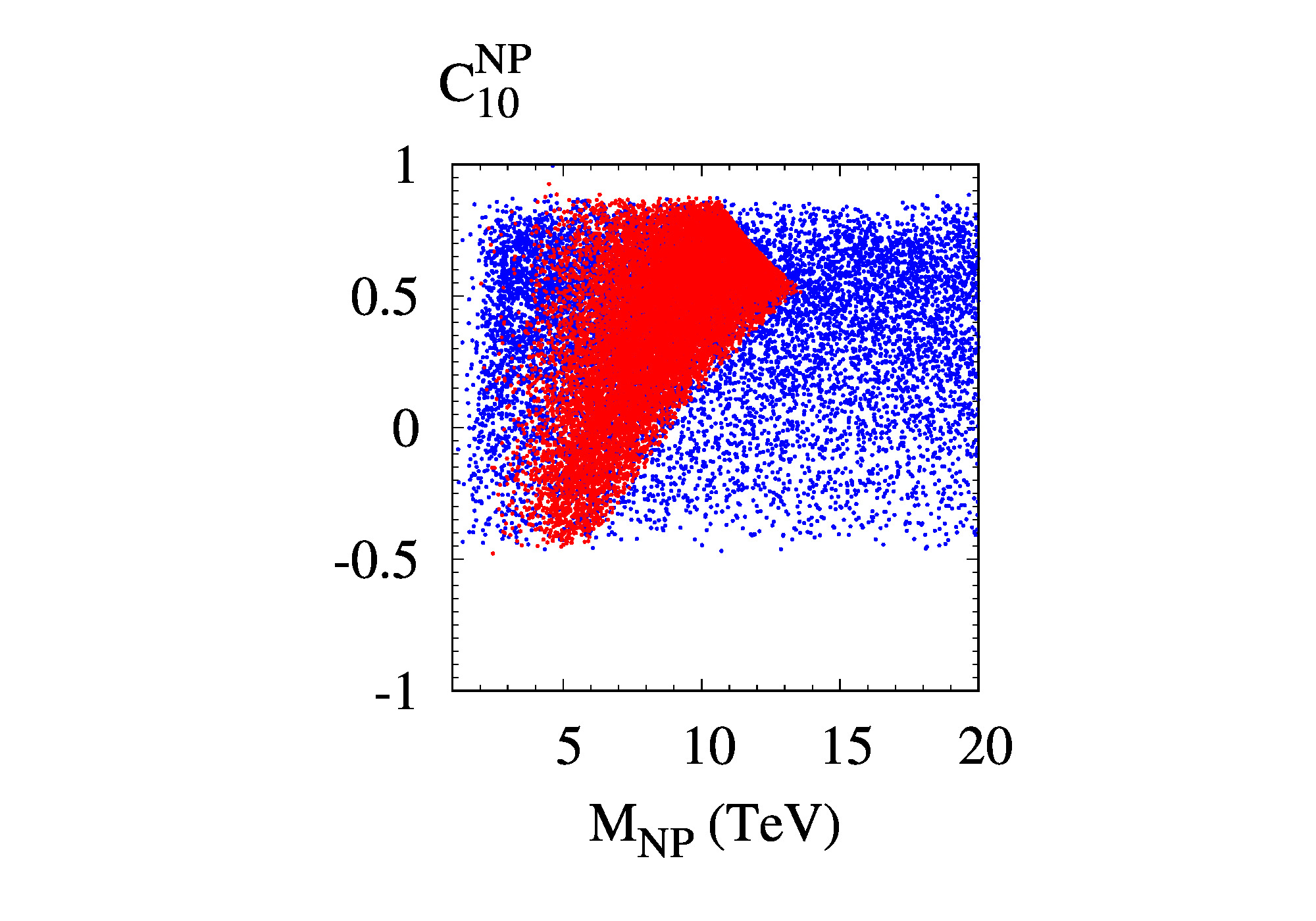} \\
(a) & (b) \\
\hspace{-1cm}\includegraphics[scale=0.12]{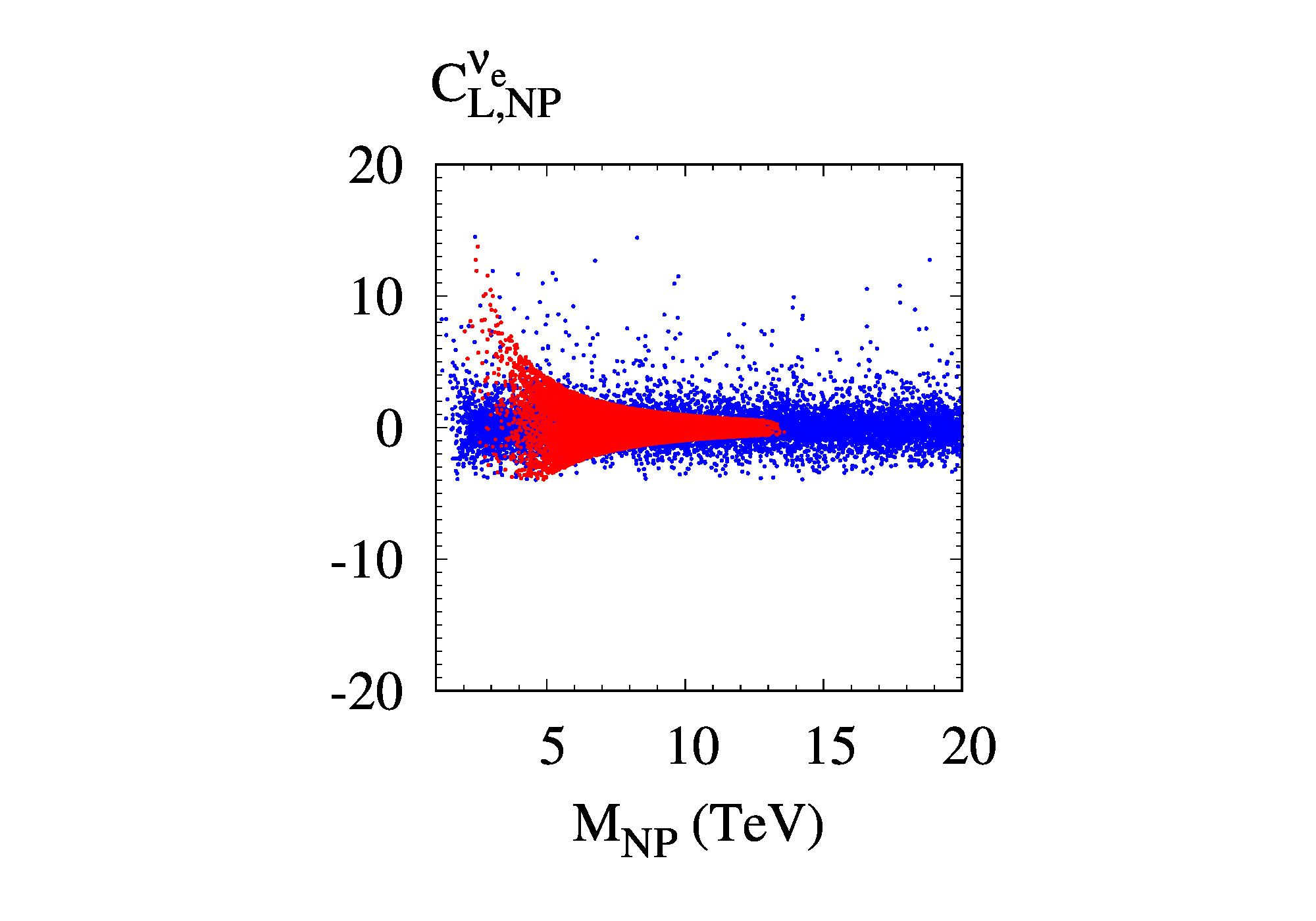} &
\hspace{-1cm}\includegraphics[scale=0.12]{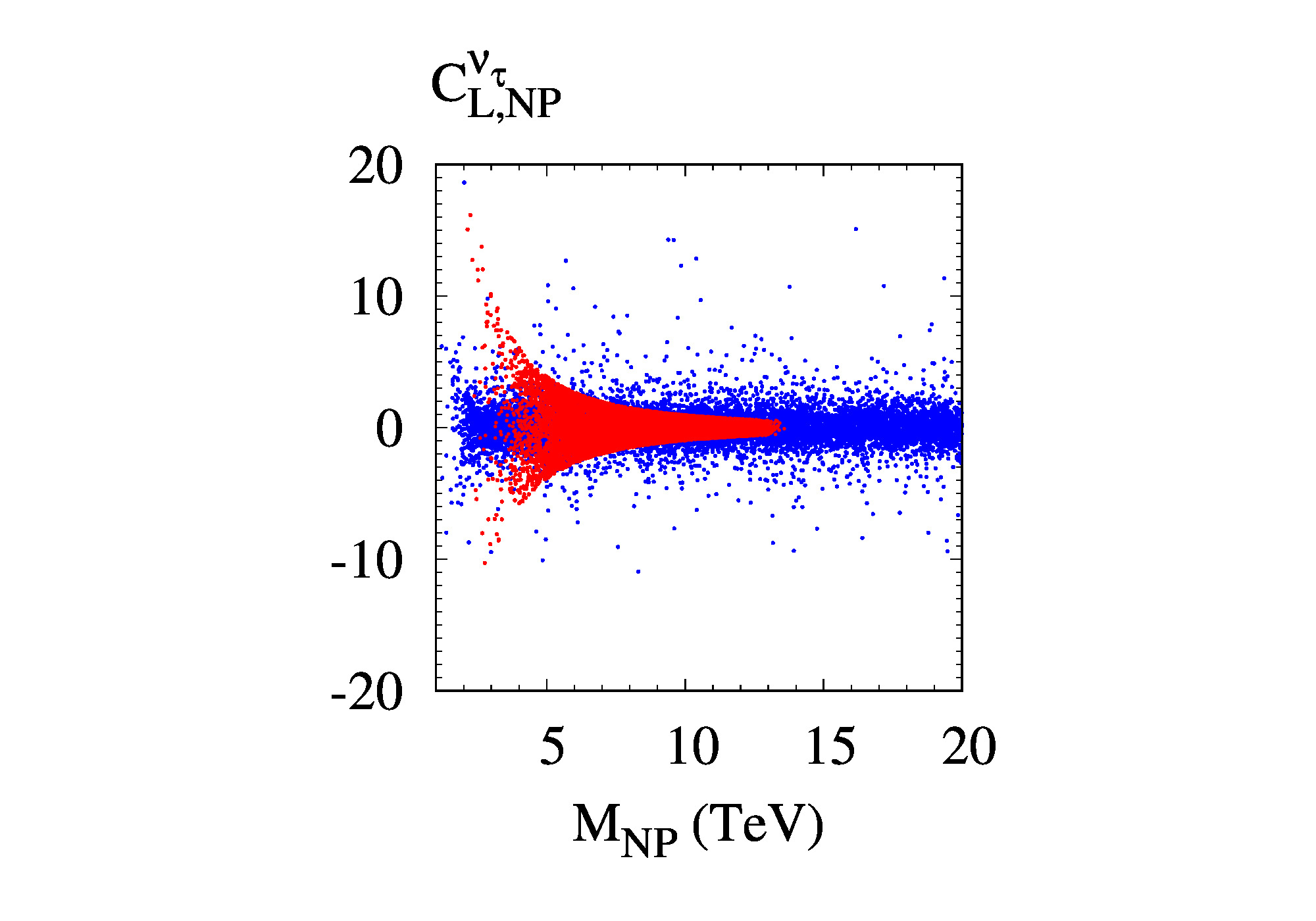} \\
(c) & (d) \\
\hspace{-1cm}\includegraphics[scale=0.12]{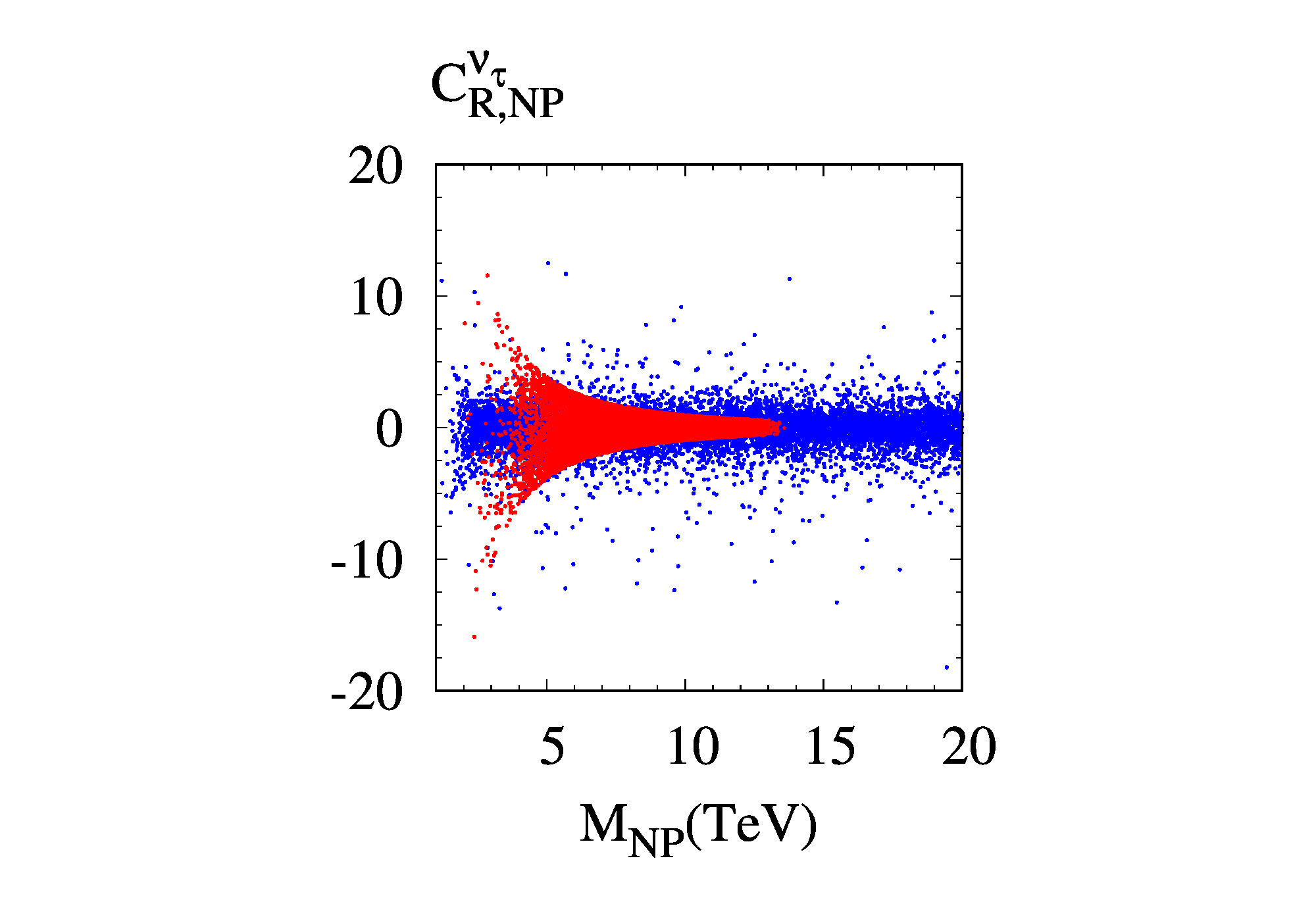} & \\
(e) &
\end{tabular}
\caption{\label{F_ap2_WC} 
Allowed regions at the $2\sigma$ level with fixed $\alpha=2$ (red dots) for
(a) $C_9^\NP$ ,  
(b) $C_{10}^\NP$, 
(c) $C_{L,\NP}^{\nu_e}$, 
(d) $C_{L,\NP}^{\nu_\tau}$, and
(e) $C_{R,\NP}^{\nu_\tau}$
vs. $M_\NP$, respectively.
Blue dots are for free $\alpha$.
}
\end{figure}
\par
Plots of Fig.\ \ref{F_ap2_WC} are for fixed $\alpha=2$ (red dots) at the $2\sigma$ level.
Red dots tend to attenuate for larger $M_\NP$ according to our parameterization 
$C_j\sim (v/M_\NP)^2$ for $\alpha=2$, 
but the specific forms are different from each other.
The case of $\alpha=2$ corresponds to ordinary heavy particle mediator.
For fixed $\alpha=2$, we find the NP window as $2.04~\TeV\le M_\NP \le 11.76~\TeV$ at the $1\sigma$ level.
The window is rather narrow compared to, for example, some of LQ masses.
According to \cite{DAlise2403}, the mass bound of a weak double scalar LQ $R_2$ is $20~\TeV$
and that of a vector weak doublet $V_2$ is $30~\TeV$ for order one couplings.
For a scalar weak triplet LQ $S_3$ and a vector weak triplet $U_3$ the masses go over about $50~\TeV$.
%
%
%
%
%
It should be remarked that the NP window as well as $R(\Lambda^{(*)})_{\nu\nu}$ are quite stable 
even though $\Br(B^+\to K^+\mu^+\mu^-)_{[0.1, 0.98]}$ is excluded in our analysis.
In this case we find that (at $1\sigma$)
$M_\NP(\alpha=2) = [2.24, 10.73]$ (TeV), 
$R(\Lambda)_{\nu\nu} = [5.80\times 10^{-3}, 3.17]$ with the best fit of $2.32$, and
$R(\Lambda^*)_{\nu\nu} = [2.36\times 10^{-3}, 1.81]$ with the best fit of $1.01$.
The $M_\NP$ window shrinks a little bit while the ranges of $R(\Lambda^{(*)})$ widen slightly
but the overall features remain unchanged, implying the robustness of our analysis.
%
%
%
%
%
\begin{figure}
\begin{tabular}{cc}
\hspace{-1cm}\includegraphics[scale=0.12]{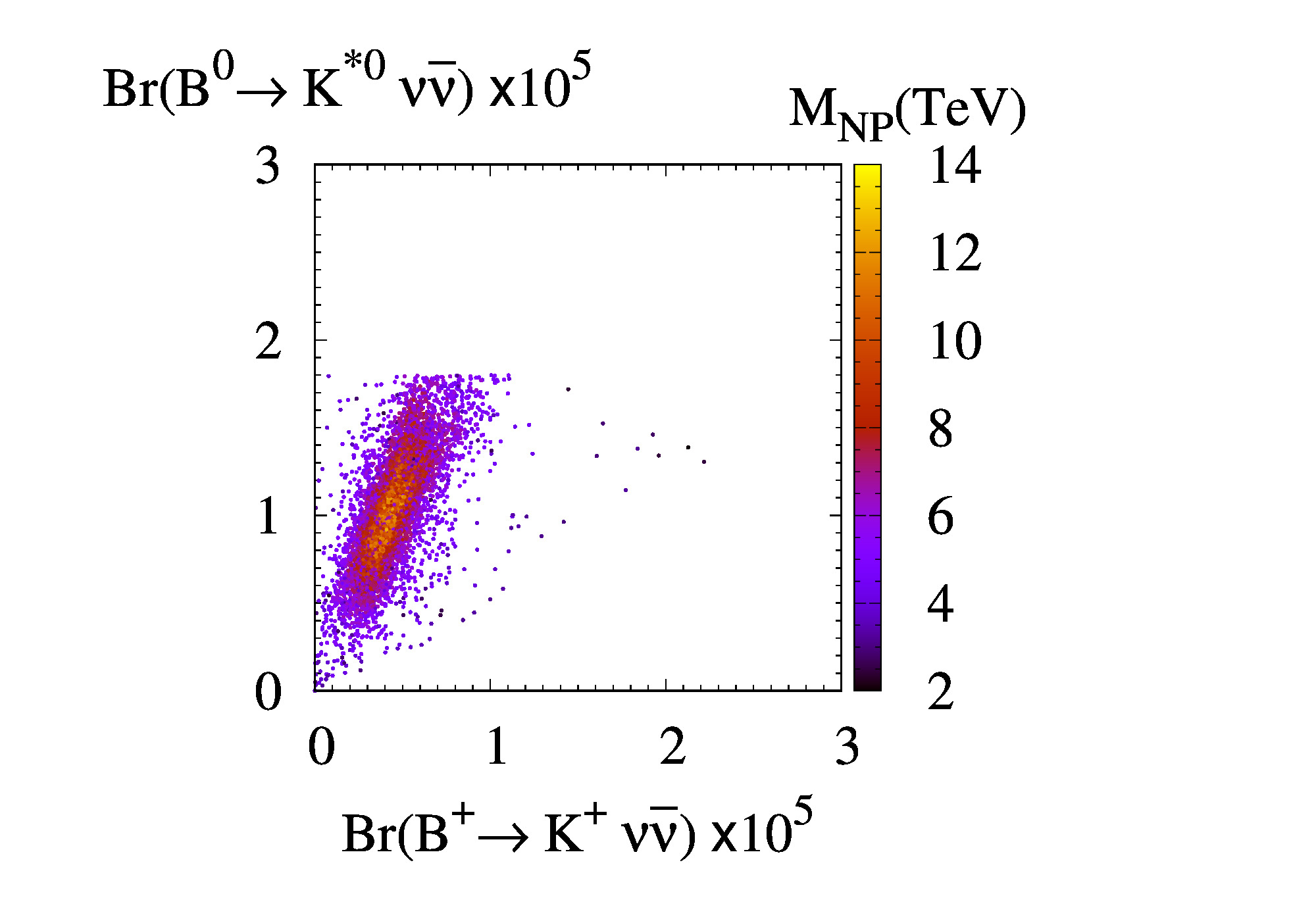} &
\hspace{-1cm}\includegraphics[scale=0.12]{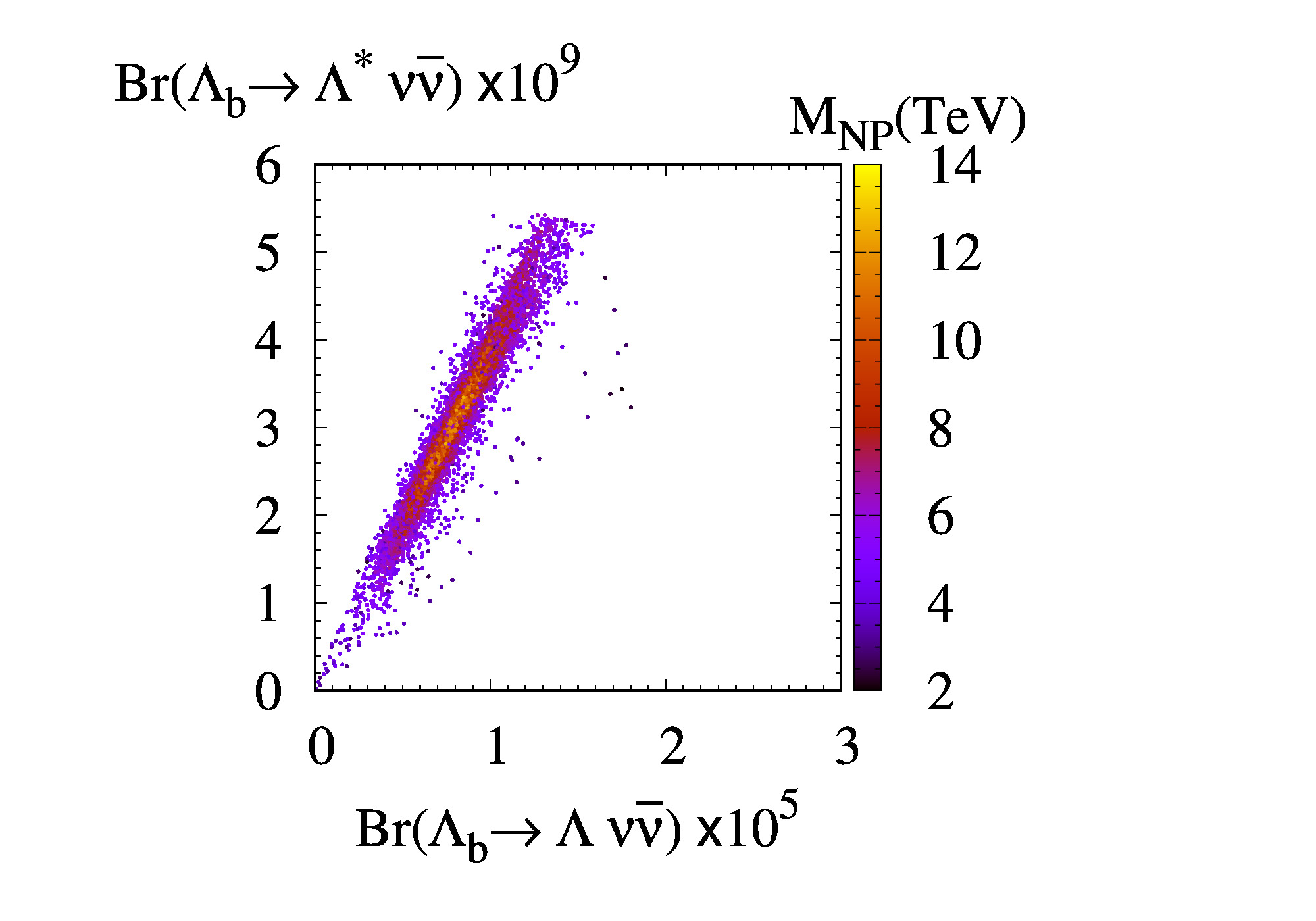} \\
(a) & (b) \\
\hspace{-1cm}\includegraphics[scale=0.12]{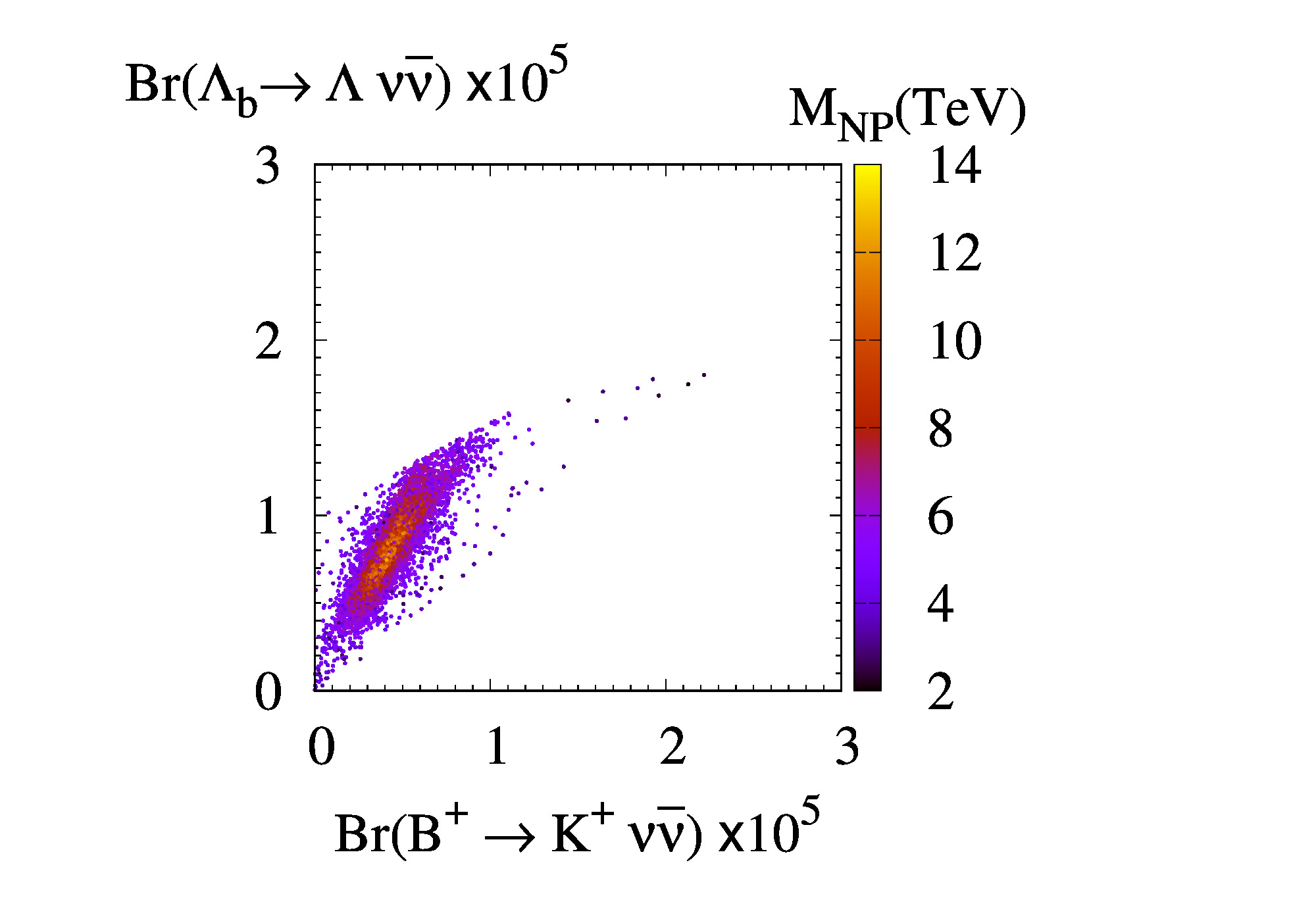} & 
\hspace{-1cm}\includegraphics[scale=0.12]{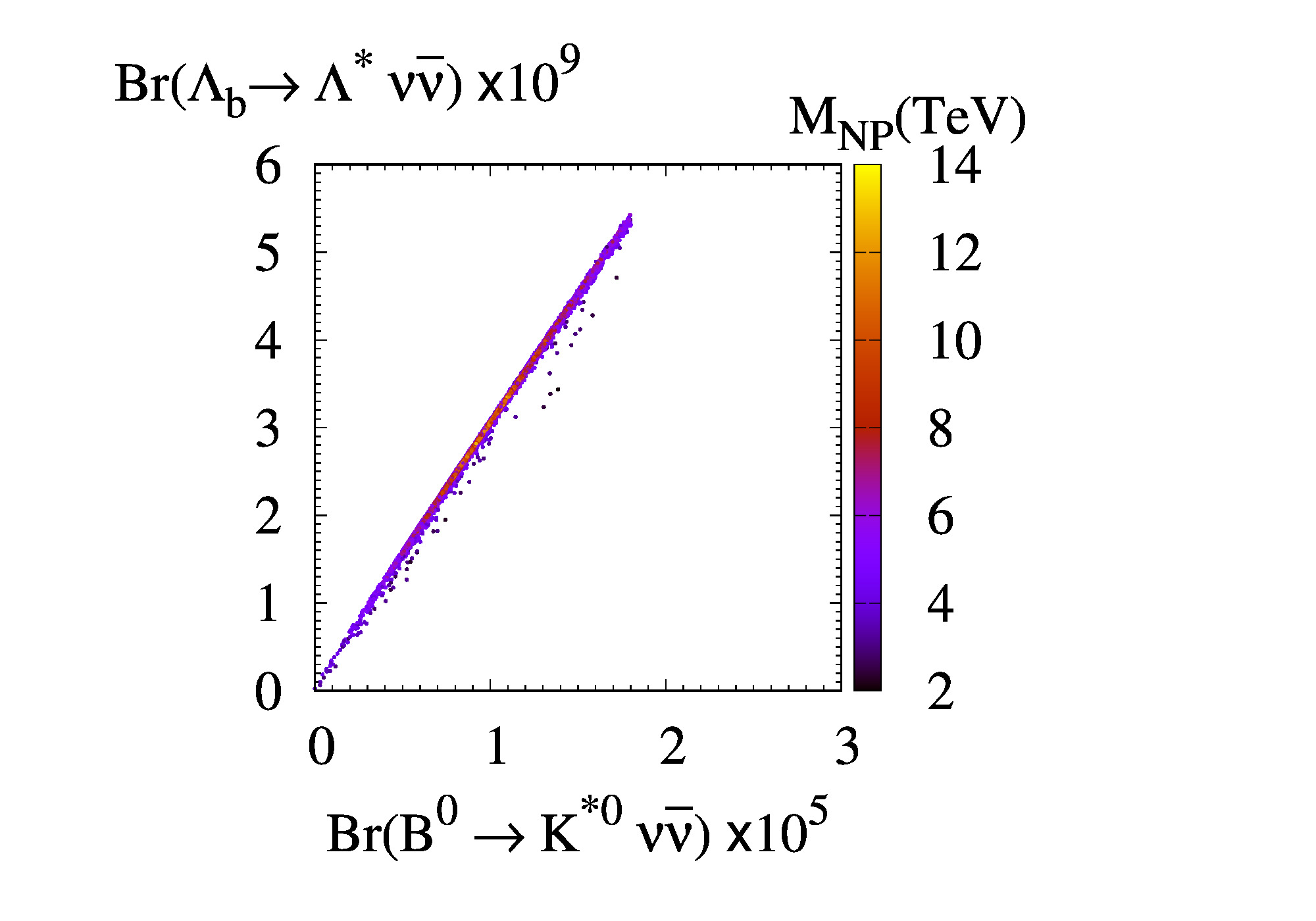}  \\
(c) & (d) \\
\hspace{-1cm}\includegraphics[scale=0.12]{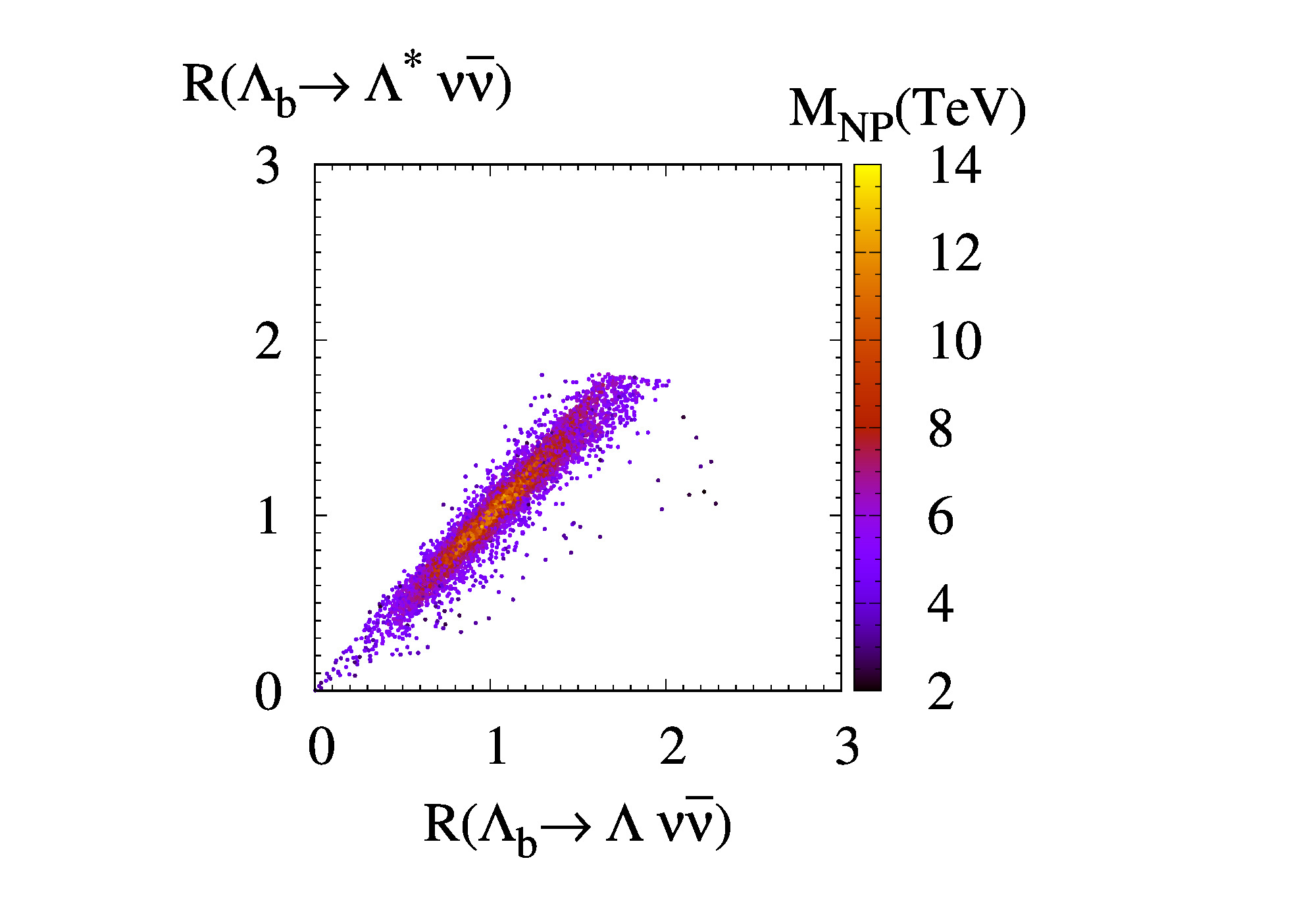} & \\
(e) &
\end{tabular}
\caption{\label{F_ap2_obs} 
Allowed regions at the $2\sigma$ level with fixed $\alpha=2$ for
(a) $\Br(B^0\to K^{*0}\nu\nubar)$ vs. $\Br(B^+\to K^+\nu\nubar)$, 
(b) $\Br(\Lambda_b\to\Lambda^*\nu\nubar)$ vs. $\Br(\Lambda_b\to\Lambda\nu\nubar)$,
(c) $\Br(\Lambda_b\to\Lambda\nu\nubar)$ vs. $\Br(B^+\to K^+\nu\nubar)$,
(d) $\Br(\Lambda_b\to\Lambda^*\nu\nubar)$ vs. $\Br(B^0\to K^{*0}\nu\nubar)$, and
(e) $R(\Lambda^*)_{\nu\nu}$ vs. $R(\Lambda)_{\nu\nu}$
with respect to $M_\NP$, respectively.
}
\end{figure}
\par
Figure \ref{F_ap2_obs} is a replica of Fig.\ \ref{F_obs} with fixed $\alpha=2$, with respect to $M_\NP$.
For higher $M_\NP$, the plots are getting closer to the SM predictions as expected.
From these Figures one can pin down the upper bound of $M_\NP$ directly by measuring the branching ratios.
\par
Experimentally, the High-Luminosity LHC (HL-LHC) of $3000~{\rm fb}^{-1}$ would reach the discovery potential
for scalar LQ mass ranges from $1.2$ to $1.7$ TeV.
The exclusion sensitivity is $1.4$ to $1.9$ TeV.
For vector LQs the exclusion sensitivity goes up to $2.8$ TeV.
For $Z'$ masses the HL-LHC could exclude as high as $6.5$ TeV \cite{ATL_22_018,CMS_22_001}.
Thus the HL-LHC is expected to probe our mass window for vector LQs and $Z'$, but our best-fit value of the
NP scale $M_{\NP,\best}=13.2$ TeV would be out of reach.
%
%
%
%
%
In addition, as mentioned in the Introduction, $\Lambda_b$ is abundantly produced at the LHCb
and $\Lambda_b \to \Lambda\mu^+\mu^-$ decay has already been observed.
For example, at the LHCb with $\sqrt{s}= 7~ \TeV$ and and integrated luminosity of $3 ~{\rm pb}^{-1}$
the production ratio of $\Lambda_b$ to $B^-$ and $\Bbar^0$, $f_{\Lambda_b}/(f_u+f_d)$
is measured to be about $\approx 25\%$ to $35\%$ with the dependence on the transverse momentum of the
charmed hadron-muon pair up to $p_T\le 14\GeV$. \cite{LHCb1111}.
The ratio of $f_{\Lambda_b}/f_d$ is observed as about $\approx 30\%$ to $69\%$ for $1.5<p_T<40 \GeV$ 
with $\sqrt{s} = 7 ~\TeV$ and an integrated luminosity of $1 ~{\rm fb}^{-1}$ \cite{LHCb1405}. 
Also future collider like FCC-ee could produce $130\times 10^9$ $\Lambda_b$ particles
\cite{Bernardi2203,Novotny2207}.
According to our results, one can expect to observe rare decay $\Lambda_b\to\Lambda^*\nu\nubar$
as well as $\Lambda_b\to\Lambda\nu\nubar$, and check compatibility with the SM.
%
%
%
%
%
\section{Conclusions}
%
In conclusion, we investigated the branching ratios of $\Br(\Lambda_b\to\Lambdas\nu\nubar)$
combined with other mesonic $b\to s$ transitions.
Relevant Wilson coefficients are parametrized as $C_j\sim(v/M_\NP)^\alpha$.
Our best fits prefer $\alpha=1.21$, suggesting that unparticle-like degrees of freedom or some other contributions
than ordinary one heavy particle are favored.
We found that the NP scale $M_\NP$ has a window of $[2.04, 11.76]~(\TeV)$ at $1\sigma$
for ordinary heavy new particles,
which is narrow compared to some of the LQ mass bounds.
The range is challenging to probe for the HL-LHC.
We predicted the branching ratios of $\Lambda_b\to\Lambdas\nu\nubar$ to be $2.07 (1.07)$ times larger
than the SM calculations.
The results are compatible with previous works but our best-fit value of 
$\Br(\Lambda_b\to\Lambda\nu\nubar)$ is rather high.
We also found that $\Br(B^0\to K^{*0}\nu\nubar)$ is expected to be close to the SM estimation.
%
%
%
%
%
Furthermore, a sum rule that interconnects mesonic and baryonic modes was explored.
It turned out that there is a very similar sum rule to that of $R_{H_c}$. 
The relation could provide complementary checks for experimental measurements.
Searching for the theoretical grounds of the sum rule would be very challenging. 
Future colliders like FCC-ee could verify our predictions.
%
%
%
%
%
%
%
%
%
%
%
\begin{acknowledgments}
This paper was supported by Konkuk University in 2024.
\end{acknowledgments}
%
%
%
%
%

\end{document}